\documentclass[article,moreauthors,pdftex,10pt,a4paper]{mdpi-arxiv}

\usepackage{amssymb,amsfonts}

\Title{Bohmian-Based Approach to Gauss-Maxwell Beams} 

\Author{\'Angel S. Sanz $^{1,}$*\orcidA{}, Milena D. Davidovi\'c $^{2}$ and~Mirjana Bo\v{z}i\'c $^{3}$}

\AuthorNames{\'Angel S. Sanz, Milena D. Davidovi\'c and~Mirjana Bo\v{z}i\'c}

\address{%
$^{1}$ \quad Department of Optics, Faculty of Physical Sciences,
Universidad Complutense de Madrid, Pza.\ Ciencias 1, Ciudad Universitaria, 28040 Madrid, Spain\\
$^{2}$ \quad Department of Mathematics, Physics and Descriptive Geometry, Faculty of Civil Engineering,
University of Belgrade, Bulevar Kralja Aleksandra 73, 11000 Belgrade, Serbia;
milena@grf.bg.ac.rs\\
$^{3}$ \quad Institute of Physics Belgrade, University of Belgrade, Pregrevica 118, 11080 Belgrade, Serbia;
bozic@ipb.ac.rs}

\corres{Correspondence: a.s.sanz@fis.ucm.es; Tel.: +34-91-394-4673}

\abstract{Usual Gaussian beams are particular scalar solutions to the~paraxial Helmholtz
equation, which neglect the~vector nature of light.
In order to overcome this inconvenience, Simon {et al.} (\emph{J.\ Opt.\ Soc.\ Am.\ A} \textbf{1986}, {\emph{3}}, 536--540)
found a~paraxial solution to Maxwell's equation in vacuum, which includes polarization in a
natural way, though still preserving the~spatial Gaussianity of the~beams.
In this regard, it seems that these solutions, known as Gauss-Maxwell beams, are~particularly appropriate
and a~natural tool in optical problems dealing with Gaussian beams acted or manipulated by polarizers.
In this work, inspired in the~Bohmian picture of quantum mechanics, a~hydrodynamic-type extension of
such a~formulation is provided and~discussed, complementing the~notion of electromagnetic field with that
of (electromagnetic) flow or streamline.
In this regard, the~method proposed has the~advantage that the~rays obtained from it render a~bona fide
description of the~spatial distribution of electromagnetic energy, since they are in compliance with the~local
space changes undergone by the~time-averaged Poynting vector.
This feature confers the~approach a~potential interest in the~analysis and~description of single-photon
experiments, because of the~direct connection between these rays and~the average flow exhibited by
swarms of identical photons (regardless of the~particular motion, if any, that these entities might have),
at least in the~case of Gaussian input beams.
In order to illustrate the~approach, here it is applied to two common scenarios, namely the~diffraction undergone
by a~single Gauss-Maxwell beam and~the interference produced by a~coherent superposition of two of such
beams.}


\newcommand{\bd}{\begin{displaymath}}
\newcommand{\ed}{\end{displaymath}}
\newcommand{\be}{\begin{equation}}
\newcommand{\ee}{\end{equation}}
\newcommand{\ba}{\begin{eqnarray}}
\newcommand{\ea}{\end{eqnarray}}

\begin{document}


\section{Introduction}
\label{sec1}

One of the~most appealing features of geometrical optics, and~also a~remarkable and~convenient advantage,
is perhaps the~fact that this model relies on the~well-defined and~very intuitive concept of ray.
This concept, which simplifies the~description and~analysis of optical processes (e.g., imaging) and~phenomena
(e.g., reflection, refraction, miracles), has a~direct physical meaning, as the~path along which the~electromagnetic
energy in the~form of light flows.
Accordingly, it has helped us to understand how light goes from one place to
another in a~simple fashion, so it is not strange that descriptions and~methodologies based on the~notion
of ray have become a~valuable tool, for instance, to infer properties of the~medium traversed by light
(e.g., refractive index, curvature, thickness, etc.) or, conversely, based on such properties, to determine
how geometrical arrangements of optical devices (e.g., lenses, mirrors, prisms, etc.) can be devised to
act on or control light.
However, when we move to the~realm of electromagnetic or wave optics, this concept blurs up.
That is, rays are still considered, but typically in a~virtual sense, as an~auxiliary tool
to evaluate relatively complex integrals, as it is the~case, for instance, to explain the
emergence of diffraction and~interference phenomena.
Accordingly, constructive and~destructive interference are associated with path
differences with a~certain value.
The~same idea, transferred to the~domain of quantum mechanics, actually
constitutes the~formal basis of the~well-known Feynman path-integral formulation
\cite{feynman-bk1}.

Such an~imbalance between geometrical and~wave optics can be overcome, though, by generalizing
the idea of physical ray to the~nonzero wavelength limit, beyond the~domain of the
geometrical optics.
A convenient starting point to this end can be a~formal extension of the~eikonal approach
to the~nonzero wavelength limit \cite{bornwolf-bk}.
Nonetheless, there is also an~alternative route, which consists in making an~effective transfer
of concepts from the~Bohmian formulation of quantum mechanics
\cite{bohm:PR:1952-1,holland-bk} to~optics.
This quantum formulation allows us to understand the~evolution of quantum systems in
terms of probability streamlines or trajectories, which denote paths along which probability
flows and~that, when are statistically considered, reproduce on a~bona-fide event-by-event
basis the~same results rendered by any other quantum formulation \cite{sanz:FrontPhys:2019}.
Although the~concept of probability is out of the~scope of classical electromagnetism (at
the level of Maxwell's equations), it is still possible to find a~beneficial correspondence
between elements from this theory and~concepts typically involved in the~Bohmian formulation.
Actually, this can readily be done if the~role of the~probability density is identified with
the electromagnetic energy density, and~the quantum density current or quantum flux
\cite{schiff-bk} with the~Poynting vector \cite{prosser:ijtp:1976-1,sanz:AnnPhysPhoton:2010,bliokh:NewJPhys:2013}.
This prescription, where the~corresponding electromagnetic streamlines or rays describe
the paths along which (electromagnetic) flows, allows to describe the~wave phenomena
accounted for by Maxwell's equations on an~event-by-event basis
\cite{sanz:AnnPhysPhoton:2010,sanz:PhysScrPhoton:2009,sanz:JRLR:2010,dimic:PhysScr:2013,davidovic:JRLR:2017}
in compliance with what one experimentally finds in low-intensity experiments
\cite{weis:AJP:2008,weis:EJP:2010}, also facilitating the~understanding
of the~statistical results typically obtained in quantum optics in the~large photon-count limit
\cite{padgett:AJP:2016} without the~need to involve Fock states in the~description.
This alternative formulation is actually not that far from the~standard formulation of
classical electromagnetism, where the~corresponding continuity equation favors the~definition
of a~velocity field relating the~electromagnetic energy density with its way to spatially distribute,
accounted for by the~time-averaged Poynting vector \cite{bornwolf-bk}.

Apart from the~intrinsic applied interest that event-by-event formulations have (analogous to
ray-tracing in the~geometrical optics limit), there are also experimental facts supporting this
view.
In~2011, Kocsis {et al.}\ reported \cite{kocsis:Science:2011} the~first reconstruction of what they
called averaged photon paths from experimental measurements of the~photon transverse momentum in a
realization of Young's two-slit experiment.
In principle, according to standard quantum arguments, this is not possible without dramatically
affecting the~interference diagram.
However, by means of a~laboratory implementation of the~idea of weak measurement
\cite{aharonov:PRL:1988,aharonov:PRA:1990,wiseman:NewJPhys:2007,dressel:RMP:2013}, it is possible to
infer information along the~transverse direction with a~slight perturbation (weak measurement) prior to the
proper detection process---what we typically regard or denote as a~strong or von Neumann measurement
\cite{vonNeumann-bk:1932En}, which gives rise to the~irreversible ``collapse'' of the~system wave function.
With the~value of the~transverse momentum at different distances from the~slits, and~taking the~Bohmian
ideas as a~basis, it was possible to reconstruct a~set of rays that were in compliance with
former calculations based on the~above prescription for electromagnetic fields
\cite{prosser:ijtp:1976-1,sanz:AnnPhysPhoton:2010}.
Thus, although the~latter is not a~probabilistic approach itself, appropriate to describe single photons
in Fock states, in the~large-number limit the~correspondence with classical electromagnetism shows there
is a~proportionality between probability distributions and~energy densities, and~therefore the
electromagnetic-flow trajectories will accurately describe the~average paths along which photons
travel.
In the~last years, the~laboratory implementation of weak measurements, used as an~alternative to quantum
tomography to determine the~photon wave function \cite{lundeen:Nature:2011,bamber:PRL:2012}, has also
inspired a~series of theoretical works concerning the~interpretation \cite{bliokh:NewJPhys:2013},
reconstruction \cite{schleich:PRA:2013}, or new observations \cite{matzkin:PRL:2012,braverman:PRL:2013}
of Bohmian trajectories.

The~above experiments typically involve Gaussian beams and, more specifically, coherent superpositions of
such beams, which include a~given polarization state.
This leads to a~series of natural questions---If a~Gaussian beam is assigned a~particular polarization state, is there
a paraxial vector description for such a~beam in the~same way there is for a~scalar one? If so, how~does the~flux
associated with such a~beam propagate along the~optical axis and~spread across the~transverse directions?
Or,~how~does the~polarization state influence the~interference between two of such beams while they evolve
along the~optical axis?
With the~purpose to provide an~answer to these questions, here we develop a~hydrodynamic description for
Gauss-Maxwell electromagnetic beams, developed in 1989 by Simon~{et~al.}~\cite{mukunda:JOSAA:1986},
in terms of electromagnetic energy flow lines or rays.
The~motivation behind this analysis is to provide a~ray-based description for localized electromagnetic
fields (Gaussian-type~beams), where both position and~polarization degrees of freedom are present, and~hence
it can readily be applied to the~analysis and~interpretation of diffraction and~interference experiments of the~kind
mentioned one.
As is well known, usual Gaussian beams are exact solutions to the~Helmholtz equation in paraxial form, but not
to Maxwell's equation under paraxial conditions.
The~approach proposed by Simon {et al.}, though, shows the~specific functional form that
the beam has to satisfy in order to be an~exact solution to Maxwell's equations in paraxial form (in vacuum).
In this regard, we would also like to mention that there are other approaches in the~literature worth exploring with
the same methodology (i.e., where the~concept of Bohmian trajectory could be exported) in order to determine the
corresponding ray equations, such as nondiffracting Helmholtz-Gauss beams \cite{bandres:JOSAA:2005}, in the~case of
scalar paraxial fields, or vector Helmholtz-Gauss beams and~related families \cite{bandres:OptLett:2005,bandres:OptExp:2006},
in~the~case of vector paraxial fields, solutions to Maxwell's equations.

In this work, in particular, we have focused on and~investigated the~formal aspects of the~rays that
describe the~spatial development of the~electromagnetic energy density in the~case of single Gauss-Maxwell
beams and~interference in the~coherent superposition of two Gauss-Maxwell beams.
Furthermore, we also consider the~limit where the~behavior of these vector beams and~their superpositions can
be described, in a~good approximation, by the~usual Gaussian beams, hence~neglecting the~associated polarization
state.
Particular interest is thus paid to the~transverse momentum, which is eventually the~observable quantity
in an~experiment and, therefore, the~quantity of interest to take the~methodology here proposed to the
analysis of real laboratory experiments.
Accordingly, this work has been organized as follows.
To be self-contained and, at the~same time, to~offer a~wider contextualization of the~work,
in Section~\ref{sec2} we introduce a~general overview of the~treatment for standard scalar Gaussian
beams, starting from a~revision of some general aspects connected to monochromatic scalar fields
in vacuum, and~how a~ray (``photon'' trajectory) equation can be properly specified for these fields.
Then, the~case of Gauss-Maxwell beams is analyzed and~discussed in Section~\ref{sec3},
first for a~linearly polarized beam and~then extended to any polarization state.
Section~\ref{sec4} is devoted to the~extension of this methodology to the~superposition
of two coherent Gauss-Maxwell beams with the~same polarization state and~also with
mutually arbitrary polarization states---a~case that may occur if a~beam is diffracted
by a~two slit and, immediately afterwards, each diffracted beam acquires a~different
polarization state.
To conclude, a~series of final remarks are summarized in Section~\ref{sec5}.


\section{Standard Scalar Gaussian Beams}
\label{sec2}

\vspace{-8pt}
\subsection{General Aspects for Monochromatic Scalar Fields}
\label{sec21}

Consider a~general electromagnetic monochromatic scalar field in vacuum \cite{bornwolf-bk},
\be
 \Xi ({\bf r},t) = \Psi({\bf r}) e^{-i\omega t} ,
 \label{general}
\ee
with wavelength $\lambda = c/\omega$.
Within a~generalized framework, the~amplitude of this field, $\Psi({\bf r})$, can be specified
as a~complex-valued static (time-independent) field satisfying Helmholtz's equation,
\be
 \nabla^2 \Psi + k^2 \Psi = 0 ,
 \label{eq1}
\ee
plus the~corresponding boundary conditions.
From this equation, the~components of the~wave vector ${\bf k} = (k_x, k_y, k_z)$ provide
us with valuable information about how the~field $\Psi({\bf r})$ changes spatially along
each space direction, while the~monochromaticity condition requires that the~wave number,
$k = \| {\bf k} \| = \sqrt{k_x^2 + k_y^2 + k_z^2} = 2\pi/\lambda$, remains constant.

Because we are dealing with light, let us assume that there is a~preferential direction
for its space propagation.
This direction is going to be referred to as the~longitudinal or parallel ($\|$) direction, which eventually
defines the~system optical axis.
Any other orthogonal direction is going to be denoted as the~perpendicular or transverse ($\perp$) direction.
Thus, in Cartesian coordinates, if we choose the~longitudinal direction along the~$z$-axis, the~$x$ and~$y$
axes specify the~transverse (mutually orthogonal) directions.
Consequently, the~position vector can be recast as ${\bf r} = (x,y,z) = ({\bf r}_\perp, z)$ and~the wave vector
as ${\bf k} = ({\bf k}_\perp, k_z)$, with the~wave number being $k = \sqrt{\|{\bf k}_\perp\|^2 + k_z^2}$.
If paraxial conditions are assumed, that is, space variations of the~field are slower along the~longitudinal direction
than along the~transverse ones, then $k_z \gg k_\perp$ and~hence $k \approx k_z$.
Accordingly, the~scalar field $\Psi({\bf r})$ can be recast as a~plane wave propagating
along the~$z$-direction, modulated by a~spatially-dependent amplitude, that is,
\be
 \Psi({\bf r}) \approx \psi({\bf r}) e^{i k_z z} \approx \psi({\bf r}) e^{i k z} .
 \label{eq2}
\ee

The~substitution of the~ansatz (\ref{eq2}) into the~Helmholtz Equation (\ref{eq1}) renders
\be
 \nabla_\perp^2 \psi + 2ik_z \frac{\partial \psi}{\partial z}
  + \frac{\partial^2 \psi}{\partial z^2} = 0 ,
 \label{eq3}
\ee
where $\nabla_\perp^2 = \partial^2/\partial x^2 + \partial^2/\partial y^2$ is the~transverse Laplacian.
Physically, paraxiality implies that the~amplitude of the~scalar field has to
smoothly change along the~longitudinal ($z$) direction.
This~allows us to neglect the~second derivative in $z$ in Equation~(\ref{eq3}), thus allowing us to simplify this
equation, which reads as
\be
 i\ \!  \frac{\partial \psi}{\partial z} = - \frac{1}{2k}\ \nabla_\perp^2 \psi .
 \label{eq4}
\ee

As it can readily be noticed, this equation is formally equivalent to the~time-dependent Schr{\H o}dinger
equation for a~free particle of mass $m$ in two dimensions \cite{sanz:JOSAA:2012}.
Actually, in this latter case, if it is assumed that propagation along $z$ is classical, we find a
relationship between this coordinate and~time:
\be
 z = \frac{p}{m}\ \! t = \frac{\hbar k}{m}\ \! t ,
 \label{eq5}
\ee
\textls[-30]{where $p = \hbar k$.
For simplicity and~convenience, we have chosen $z_0 = 0$ in (\ref{eq5}), although this is not necessary.}

It is worth highlighting that, in spite of its apparent simplicity, Equation~(\ref{eq5}) brings in
two important consequences at a~conceptual level:
\begin{enumerate}[leftmargin=*,labelsep=5mm]
 \item This relation enables a~direct switch from propagation in time of an~extended wave, namely
 the~scalar field $\Psi({\bf r})$, to space diffusion along the~longitudinal (axial) direction of a~particular
 ``slice'' of such a~wave.
 More specifically, if we consider a~transverse section or plane of the~full wave [consider, for instance, that such
 a~wave describes a~pulse with amplitude $\psi({\bf r})$] within the~$XY$ plane for a~given value $z_0$ (in other words,
 the~input plane $z = z_0$, analogous to considering $t_0$ in a~time-propagation), we shall obtain its spatial redistribution
 or accommodation to the~corresponding boundaries at subsequent planes, with $z$ increasing.
 Consequently, if~the~pulse has an~extension along the~$z$ direction, considering different ``slices'' of the~pulse (i.e.,~different~$z_0$~planes), we may easily determine the~shape of the~pulse at a~further distance by just combining all the
 resulting ``slices'' $z_f$.
 Notice that this fact also allows to establish the~validity limit for the~approximation, which is going to remain correct
 provided dispersion along the~$z$ direction can be assumed to be negligible (i.e., as long as all ``slices'' travel with
 nearly the~same speed).

 \item On the~other hand, such a~relation enables a~simple, direct link between wave optics and~matter-wave optics,
 which arises from the~formal relation established between Schr{\H o}dinger's equation and~the Helmholtz equation
 in paraxial form, both being parabolic differential equations describing the~transport of the~quantity
 $\Psi$ (regardless of the~nature of such a~quantity, that is, whether it describes an~electromagnetic amplitude or
 a~probability amplitude) with an~imaginary diffusion constant (this complex valuedness is precisely the~fundamental
 trait that allows interference in the~solution of these equations, but not in the~heat equation, although it is also
 a~differential equation of the~parabolic class).
 This is also a~rather convenient issue both analytically and~computationally, because it explicitly shows that
 optical and~matter-wave problems ruled by the~same equation form have the~same solution and, eventually,
 the~same interpretations \cite{sanz:neutron:2020}.
\end{enumerate}

Taking into account point~2 above, Equation~(\ref{eq4}) can readily be recast in the~form of a
Schr{\H o}dinger-type equation by replacing the~coordinate $z$ with the~value indicated by the~relation~(\ref{eq5})---notice, however, that this gives rise to a~two-dimensional Schr{\H o}dinger equation, since the~only space coordinates,
after substitution, are $x$ and~$y$.
This direct analogy can be taken a~step further to introduce a~guidance equation in the~Bohmian form, namely
\be
 \frac{d{\bf r}_\perp}{dz} = \frac{\nabla_\perp S_\psi}{k} ,
 \label{eq6}
\ee
with
\be
 S_\psi = \frac{1}{2i}\ \! \ln \left(\frac{\psi}{\psi^*}\right)
\ee
describing the~space phase variations of the~complex field amplitude $\psi$.
Since $d{\bf r}'_\perp/dz \approx {\bf k}_\perp/k$, this equation allows us to describe the~transverse distribution
of electromagnetic energy in terms of streamlines by providing the~corresponding initial conditions and~integrating
it along $z$ \cite{sanz:JOSAA:2012}.
Further, observe that also in this case, the~connection between Equation~(\ref{eq6}) and~the Bohmian equation for
matter waves is straightforward, since the~latter can be directly obtained by considering the~relation (\ref{eq5})
and the~change $S_\psi \to S/\hbar$, from the~phase of the~amplitude $\psi$ to the~phase of the~matter wave
(with~$\hbar$~emphasizing the~fact that $S$ has units of action).


\subsection{Gaussian Beam Propagation}
\label{sec31}

Consider that at the~input plane $z_0=0$, the~electromagnetic field is described by a
beam with Gaussian amplitude (on the~$z_0$ plane),
\be
 \psi({\bf r}_\perp;0) = \frac{1}{\sqrt{2\pi\sigma_0^2}}
   e^{-\|{\bf r}_\perp - {\bf r}_{\perp,c}\|^2/4\sigma_0^2} ,
 \label{eq13}
\ee
which may represent a~Gaussian mode released from an~optical fiber or
the light coming out from a~laser pointer.
The~beam (\ref{eq13}) is centered at ${\bf r}_{\perp,c}=(x_c,y_c)$ and~its width is related to its waist,
$w_0$, by the~simple relation $\sigma_0 = w_0/2$.
In this latter regard, we take here the~conventional definition for the~waist of a~Gaussian beam, as its
size at the~point of its focus (here located at the~input plane, $z_0=0$), which~corresponds to the~radius
of the~$1/e^2$ irradiance contour at the~plane ($z_0$) where the~wavefront is flat.
In order to provide some typical values, we can take them from the~experiment reported in Reference \cite{kocsis:Science:2011}.
In~this experiment, a~coherent superposition of two nearly Gaussian beams \mbox{(in a~good approximation \cite{dimic:PhysScr:2013})}
is~generated, with waists $w_0 = 0.608$~mm, wavelength $\lambda = 943$~nm, and~their centers separated a~distance
$d = 4.69$~mm.

The~propagation of the~amplitude (\ref{eq13}) along the~$z$-axis is obtained
by acting on it with the~free-space propagator $\hat{\mathcal{U}}$, that is,
\be
 \psi({\bf r}_\perp;z) = \hat{\mathcal{U}}(\hat{\bf r}_\perp,\hat{\bf p}_\perp) \psi({\bf r}_\perp;0) ,
\ee
which is equivalent to considering the~integral
%
\be
 \psi({\bf r}_\perp;z) = \frac{1}{i\lambda z}
  \int \psi({\bf r}'_\perp,0) e^{ik \| {\bf r}_\perp - {\bf r}'_\perp \|^2/2z} d{\bf r}'_\perp
 \label{eq14}
\ee
%
(for a~derivation of this expression, see Appendix~\ref{appA}).
Accordingly, the~substitution of the~ansatz (\ref{eq13}) into the~integral (\ref{eq14}) renders
\be
 \psi({\bf r}_\perp;z) = A_z
         e^{-\|{\bf r}_\perp - {\bf r}_{\perp,c}\|^2/4\sigma_0\tilde{\sigma}_z} ,
 \label{eq15}
\ee
where
\be
 A_z = \frac{1}{\sqrt{2\pi\tilde{\sigma}_z^2}} = \frac{e^{-i\varphi_z}}{\sqrt{2\pi\sigma_z^2}}
\ee
\textls[-25]{is a~complex-valued norm factor, $\varphi_z$ is the~well-known Gouy phase in optics
(typical of Gaussian beams),}
\be
 \varphi_z = (\tan)^{-1} \left( \frac{z}{2k\sigma_0^2} \right) ,
\ee
and
\be
 \tilde{\sigma}_z = \sigma_0 \left[1 + \frac{iz}{2k\sigma_0^2} \right]
\ee
is an~also complex-valued spread factor.
The~dispersion of the~beam at a~distance $z$ from the~slits is given by the~expression
\be
 \sigma_z = |\tilde{\sigma}_z| = \sigma_0 \sqrt{1 + \left( \frac{z}{2k\sigma_0^2} \right)^2} ,
 \label{sigmaz}
\ee
related, by means of the~simple relation $\sigma_z = w_z/2$, to radius of the~$1/e^2$ contour
when the~wave has propagated a~distance $z$,
\be
 w_z = w_0 \sqrt{ 1 + \left( \frac{2z}{kw_0^2} \right)^2} .
 \label{eq16}
\ee

As it can be noticed from (\ref{sigmaz}), the~expansion of the~beam along the~optical axis can be
characterized in terms of the~Rayleigh length,
\be
 z_R = 2k\sigma_0^2 .
 \label{critic}
\ee
which constitutes a~critical or characteristic length that sets a~separation between two dynamically
different regimes regarding the~beam expansion, namely the~Fresnel or hyperbolic expansion regime
and the~Fraunhofer or linear expansion regime \cite{sanz:AJP:2012}.

The~intensity or irradiance obtained from (\ref{eq15}) is
\be
 I({\bf r}_\perp;z) = |\psi({\bf r}_\perp;z)|^2 = \frac{1}{2\pi\sigma_z^2}\ \! e^{-\|{\bf r}_\perp - {\bf r}_{\perp,c}\|^2/2\sigma_z^2} .
 \label{eq17}
\ee

The~streamlines that describe the~spatial propagation of this intensity along
$z$ can be obtained, according to Equation~(\ref{eq6}), from the~equation of motion
\be
 {\bf r}'_\perp = \frac{\nabla_\perp S_\psi}{k} = \frac{1}{k}\ {\rm Im} \left[ \frac{\nabla_\perp \psi}{\psi} \right]
  = \frac{1}{2ikI} \left[ \psi^* \nabla_\perp \psi - \psi \nabla_\perp \psi^* \right] ,
\ee
which, particularized to each transverse coordinate, leads to
%
\ba
 \frac{dx}{dz} & = & \frac{1}{k}\frac{\partial S_\psi}{\partial x}
  = \frac{z}{4k^2\sigma_0^2}\frac{(x - x_c)}{\sigma_z^2} ,
 \label{eq18} \\
 \frac{dy}{dz} & = & \frac{1}{k}\frac{\partial S_\psi}{\partial y}
  = \frac{z}{4k^2\sigma_0^2}\frac{(y - y_c)}{\sigma_z^2} .
 \label{eq19}
\ea
%

The~integration of Equations~(\ref{eq18}) and~(\ref{eq19}) is straightforward, leading to
%
\ba
 x_z & = & x_c + \frac{\sigma_z}{\sigma_0} \left( x_0 - x_c \right) ,
 \label{eq20} \\
 y_z & = & y_c + \frac{\sigma_z}{\sigma_0} \left( y_0 - y_c \right) ,
 \label{eq21}
\ea
%
where $x_0$ and~$y_0$ denote the~position of the~optical streamline at the~input plane $z=0$,
while its position on the~output plane $z$ is given by $x_z$ and~$y_z$.

Notice that the~solutions rendered by Equations~(\ref{eq20}) and~(\ref{eq21}) display radial symmetry with respect to
the center of the~Gaussian beam, $(x_c,y_c)$.
Accordingly, if their initial positions distribute around this point in a~circle of radius $\rho_0 = \|{\bf r}_{\perp,0} - {\bf r}_{\perp,c}\|$, they
will distribute on a~plane at a~distance $z$ around a~circle of radius $\rho_z = \|{\bf r}_{\perp,z} - {\bf r}_{\perp,c}\|$ that has increased its
size in a~factor $\sigma_z/\sigma_0$ with respect to~$\rho_0$.
This is better seen if, from the~definitions of $\rho_0$ and~$\rho_z$, the~two equations are rewritten in a~more compact form as:
\be
 \rho_z = \frac{\sigma_z}{\sigma_0}\ \! \rho_0 .
 \label{rho}
\ee

Relatively close to the~input plane, this increase is negligible and~then, as $z$ starts becoming
important compared to the~critical value $z_R$, there is a~quadratic dependence
on $z$, as expected in a~typical Fresnel regime.
In contrast, asymptotically, far from the~input plane, the~increase of the~radius is linear with $z$,
in agreement with a~propagation in a~Fraunhofer regime.
This can also be noticed from the~expression describing the~expansion rate of the~beam,
\be
 \frac{1}{\rho_z} \frac{d\rho_z}{dz} = \frac{z}{z^2 + z_R^2} ,
 \label{expand}
\ee
which can be obtained from Equations~(\ref{eq18}) and~(\ref{eq19}) [or also directly from (\ref{rho})]
and describes the~beam expansion in relation to the~actual beam size.
This expression displays a~maximum for $z=z_R$.
Thus, before reaching the~position of this critical plane, the~expansion rate is first negligible
(for $z \lll z_R$) and~then starts increasing nearly linear with $z$ (for $z \ll z_R$).
This implies that the~radius will undergo a~sort of initial fast boost that will accelerate its increase
and then the~expansion of the~beam.
However, after surpassing the~critical plane, the~rate falls with $z$, which will keep the~radius increasing
at a~constant rate, and~hence the~divergence displayed by any ray with respect to the~center of
the input beam, consistent with the~fact that far from the~input plane ($z \to \infty$), the~radial speed,
\be
  \frac{d\rho_z}{dz} = \frac{z}{z_R \sqrt{z^2 + z^2_R}}\ \! \rho_0 ,
\ee
becomes a~constant.
This is in compliance with the~linear increase of the~cross section that we observe in Gaussian beams far from
the source (e.g., a~non-collimated laser beam).

\textls[-5]{Furthermore, it is also worth noting that the~expression for the~expansion rate (\ref{expand}) is
also the~same as the~wavefront curvature of the~Gaussian beam, with its inverse providing the~radius
of curvature of the~latter.
In this regard, we have that wavefronts are flat (infinite radius of curvature) at the~input
plane $z=0$ (i.e., at the~beam waist) and~also as $z$ becomes large compared to $z_R$ (i.e.,
in the~Fraunhofer~regime).
On the~contrary, they undergo maximal curvature when $z = z_R$ (with the~radius of curvature
being~$2z_R$), just at the~plane where the~beam expansion rate (\ref{expand}) gets its maximum
value, $1/2z_R$.}


\section{Gauss-Maxwell Beams}
\label{sec3}

\vspace{-8pt}
\subsection{General Considerations on the~Propagation Procedure}
\label{sec22}

Gaussian beams are exact solutions to the~paraxial Equation (\ref{eq4}), as seen above.
They are typically interpreted as the~amplitude associated with electric and~magnetic field vectors,
transverse to the~beam axis at any $z$ value and~everywhere polarized in the~same direction.
However, they are not exact paraxial solutions to Maxwell's equations.
Solutions that satisfy this requirement were found by \mbox{Simon {et al.}\ \cite{mukunda:JOSAA:1986},}
who called them Gauss-Maxwell beams (for other alternative but equivalent procedures, \mbox{see
References~\cite{davis:PRA:1979,mcdonald:notes:2000,mcdonald:notes:2009-1,mcdonald:notes:2009-2}).}
These beams can be determined according to the~prescription that we shall describe now,
and that will be considered from now on.

Consider the~general form of a~scalar solution to the~paraxial Helmholtz equation, Equation~(\ref{eq4}),
\be
 \psi({\bf r}_\perp;z) = \hat{\mathcal{U}}(\hat{\bf r}_\perp,\hat{\bf p}_\perp)
   \psi({\bf r}_\perp;0) ,
 \label{eq7}
\ee
where
\be
 \hat{\mathcal{U}}(\hat{\bf r}_\perp,\hat{\bf p}_\perp) = e^{iz\hat{\bf p}_\perp^2/2k}
 \label{eq7b}
\ee
is the~free-space propagator, with $\hat{\bf r}_\perp = {\bf r}_\perp$ and
$\hat{\bf p}_\perp = -i\nabla_\perp$ denoting the~transverse position and~momentum
operators, respectively.
According to Equation~(\ref{eq7}), the~spatial distribution of electromagnetic energy can be easily
determined by considering an~input beam $\psi({\bf r}_\perp;0)$ at $z=0$ and~then evolving
it along $z$ by means of the~propagator (\ref{eq7b}).
According to the~discussion in the~previous section, this propagator provides us with the~transverse
distribution of electromagnetic energy at each $z$-plane.
Although this procedure is for scalar fields, a~similar procedure can also be followed for the~vector fields governed
by Maxwell's equations by recasting the~electromagnetic field as a~six-component vector field (to some extent,
analogous to the~so-called Riemann-Silberstein electromagnetic vector \cite{sanz:AnnPhysPhoton:2010}),
\be
 {\bf F}({\bf r}_\perp;z) = \frac{1}{\sqrt{2}} \left( \begin{array}{c}
   \sqrt{\epsilon_0} E_x({\bf r}_\perp;z) \\
   \sqrt{\epsilon_0} E_y({\bf r}_\perp;z) \\
   \sqrt{\epsilon_0} E_z({\bf r}_\perp;z) \\
   \sqrt{\mu_0} H_x({\bf r}_\perp;z) \\
   \sqrt{\mu_0} H_y({\bf r}_\perp;z) \\
   \sqrt{\mu_0} H_z({\bf r}_\perp;z) \end{array} \right) ,
 \label{eq8}
\ee
where ${\bf H} = {\bf B}/\mu_0$ is used instead of ${\bf B}$ for simplify, as it will be seen below.

The~vector field ${\bf F}({\bf r}_\perp;z)$ at the~output plane $z$ arises after propagating a~distance $z$ an~input
vector field ${\bf F}({\bf r}_\perp;0)$, with the~evolution being described by a~certain operator that has to be
determined.
To this end, notice that if the~evolution of an~input scalar field is described by the~propagator~(\ref{eq7b}),
then the~evolution of ${\bf F}$ should be described by a~6$\times$6-matrix operator, henceforth denoted by~$\hat{\underline{\mathbb{U}}}$.
This~operator is obtained by replacing the~transverse position vector ${\bf r}_\perp = (x,y)$ in the~scalar operator~(\ref{eq7b}) by a~more general position vector operator \cite{mukunda:JOSAA:1986}, with matrix elements given by
\be
 \mathbb{R}_\perp = {\bf r}_\perp \mathbb{I} + k^{-1} \mathbb{G}_\perp
 = (x\mathbb{I} + k^{-1} \mathbb{G}_x,\ y\mathbb{I} + k^{-1} \mathbb{G}_y) .
\ee

In other words, we have a~transformation
\be
 \hat{\mathcal{U}}(\hat{\bf r}_\perp,\hat{\bf p}_\perp)\ \! \psi({\bf r}_\perp;0) \longrightarrow
  \hat{\underline{\mathbb{U}}}(\hat{\bf r}_\perp \mathbb{I} + k^{-1}\mathbb{G}_\perp, \hat{\bf p}_\perp\mathbb{I})\ \! {\bf F}({\bf r}_\perp;0) ,
 \label{eq9}
\ee
where $\mathbb{I}$ is a~6$\times$6 identity matrix, and~the $\mathbb{G}$-matrices are defined as
\be
 \mathbb{G}_x = \frac{1}{2} \left( \begin{array}{cc}
  - \mathbb{S}_2 & \mathbb{S}_1 \\
  - \mathcal{S}_1 & - \mathbb{S}_2 \end{array} \right) , \qquad
 \mathbb{G}_y = \frac{1}{2} \left( \begin{array}{cc}
  \mathbb{S}_1 & \mathbb{S}_2 \\
  - \mathbb{S}_2 & \mathbb{S}_1 \end{array} \right) ,
 \label{eq10}
\ee
with
\be
 \mathbb{S}_1 = \left( \begin{array}{ccc}
  0 & 0 & 0 \\ 0 & 0 & - i \\ 0 & i & 0 \end{array} \right) , \qquad
 \mathbb{S}_2 = \left( \begin{array}{ccc}
  0 & 0 & i \\ 0 & 0 & 0 \\ - i & 0 & 0 \end{array} \right) .
 \label{eq11}
\ee

The~$\mathbb{G}$-matrices satisfy the~properties:
%
\ba
 [ \mathbb{G}_a, \mathbb{G}_b] & = & 0 ,
 \label{eq23} \\
 \sum_a \mathbb{G}_a \mathbb{G}_a & = & 0 ,
 \label{eq24} \\
 \mathbb{G}_a \mathbb{G}_b \mathbb{G}_c & = & 0 ,
 \label{eq24b}
\ea
%
\noindent
for $a,b,c = x,y$, which can be proven taking into account the~following matrix relations:
\ba
 \mathbb{S}_1^{2n} & = & \left( \begin{array}{ccc}
  0 & 0 & 0 \\ 0 & 1 & 0 \\ 0 & 0 & 1 \end{array} \right) = \mathbb{S}_1^2 , \qquad
 \mathbb{S}_1^{2n-1} = \mathbb{S}_1 , \\
 \mathbb{S}_2^{2n} & = & \left( \begin{array}{ccc}
  1 & 0 & 0 \\ 0 & 0 & 0 \\ 0 & 0 & 1 \end{array} \right) = \mathbb{S}_2^2 , \qquad
 \mathbb{S}_2^{2n-1} = \mathbb{S}_2 , \\
 & & \mathbb{S}_2^2 - \mathbb{S}_1^2 = \left( \begin{array}{ccc}
  1 & 0 & 0 \\ 0 & -1 & 0 \\ 0 & 0 & 0 \end{array} \right) , \\
 & & \left[ \mathbb{S}_1, \mathbb{S}_2 \right]_+ =\left( \begin{array}{ccc}
  0 & -1 & 0 \\ -1 & 0 & 0 \\ 0 & 0 & 0 \end{array} \right) , \\
 & & \left[ \mathbb{S}_1, \left[ \mathbb{S}_1,\mathbb{S}_2 \right]_+ \right]_+ = \mathbb{S}_2 , \\
 & & \left[ \mathbb{S}_2, \left[ \mathbb{S}_2,\mathbb{S}_1 \right]_+ \right]_+ = \mathbb{S}_1 , \\
 & & \mathbb{S}_1 \mathbb{S}_2 \mathbb{S}_1 = \mathbb{S}_2 \mathbb{S}_1 \mathbb{S}_2 = 0 ,
\ea
with $n \ge 1$ and~where $[\cdot,\cdot{]}_+$ denotes the~anticommutator.
Note that the~action of $\underline{\hat{\mathbb{U}}}$ on the~vector field ${\bf F}$ can be accomplished
in two steps:
\begin{enumerate}[leftmargin=*,labelsep=5mm]
 \item \textls[-5]{First, there is a~space translation, from ${\bf r}_\perp \mathbb{I}$ to $\mathbb{R}_\perp$, by an~effective
 amount $k^{-1} \mathbb{G}_\perp$, proportional to $\lambda$.}

 \item Then, the~beam undergoes a~boost, accounted for by the~action of the~momentum operator $\hat{\bf p}_\perp\mathbb{I}$,
  while it is freely propagating along the~$z$-direction.
\end{enumerate}

The~above two-step prescription allows us to determine in a~relatively simple fashion the~evolution (along $z$) of Gauss-Maxwell beams
(or any linear combination of them) once the~input amplitude, ${\bf F}({\bf r}_\perp;0)$, is known.
Monitoring this evolution with rays, which are in compliance with the~paraxial form of Maxwell's equations,
can be done now with the~aid of the~time-averaged Poynting vector, since
\be
 {\bf S} = \langle \mathcal{E}({\bf r},t) \times \mathcal{H}({\bf r},t) \rangle_T
  = \frac{1}{2}\ \! {\rm Re} \left\{ {\bf E} \times {\bf H}^* \right\} ,
 \label{avS}
\ee
where $\mathcal{E}({\bf r},t)$ and~$\mathcal{H}({\bf r},t)$ denote, respectively, the~electric and~magnetic monochromatic
vector fields, solutions to the~Maxwell equations, and~$\langle \phantom{o} \phantom{o} \rangle_T$ is the~average over
the period of the~radiation.
Notice that this averaging allows us to also recast the~time-averaged Poynting vector just in terms of the~time-independent amplitudes
${\bf E}({\bf r})$ and~${\bf H}({\bf r})$, which arise from the~generalization of $\mathcal{E}({\bf r},t)$ and~$\mathcal{H}({\bf r},t)$,
respectively, to the~complex domain \cite{bornwolf-bk}, but that here are directly determined from the~output beam
${\bf F}({\bf r}_\perp;z)$, at a~distance $z$ from the~input (transverse) plane.
In this case, the~guidance equation is defined as
\be
 \frac{d{\bf r}_\perp}{dz} = \frac{\bf S_\perp}{S_z} ,
 \label{eq12}
\ee
where  ${\bf S}_\perp$ and~$S_z$ are, respectively, the~transverse and~longitudinal components of the
time-averaged Poynting vector.
As before, Equation~(\ref{eq12}) corresponds to a~phase velocity, with its validity being determined by the~fact that
we are dealing with vacuum, where the~phase velocity and~the components of the~time-averaged Poynting vector are proportional
(in general, for nondispersive media, phase velocity and~group velocity point in the~same direction \cite{bornwolf-bk}).


\subsection{Linearly Polarized Gauss-Maxwell Beams}
\label{sec32}

As seen above, the~ray ${\bf r}_{\perp,z} =(x_z,y_z)$ described by Equations~(\ref{eq20}) and~(\ref{eq21})
allows to understand the~distribution of the~electromagnetic energy along the~$z$ axis when the~latter
is specified by only a~scalar field.
In this regard, such rays have analogous properties to those that we find for quantum wave
packets \cite{sanz:cpl:2007,sanz:AJP:2012}.
However, they do not contain any vector-type information, because they have not been obtained from
the paraxial form of Maxwell's equations, but from the~paraxial approximation applied to Helmholtz's
equation.
To this end, first we need to determine the~electric and~magnetic field
components associated with the~evolution of the~input Gaussian amplitude (\ref{eq13}), which is done
with the~aid of the~transformation relation (\ref{eq9}), and~then the~rays are determined from
Equation~(\ref{eq12}).
In this latter regard, it is interesting to compare the~outcome from this equation for a~Gauss-Maxwell beam
with the~rays described by Equations~(\ref{eq20}) and~(\ref{eq21}) for a~bare Gaussian beam.

For simplicity in the~analysis, here we are going to consider the~case of horizontal polarization (along
the $x$ axis; vertical polarization is taken along the~$y$-axis), leaving the~case of arbitrary polarization
for the~next section.
Accordingly, the~input electromagnetic vector field, ${\bf F}({\bf r}_\perp;0)$, reads as
\be
 {\bf F}({\bf r}_\perp;0)
  = \frac{1}{\sqrt{2}}\ \psi({\bf r}_\perp;0) \left( \begin{array}{c}
   \sqrt{\epsilon_0} E_0 \\ 0 \\ 0 \\ 0 \\ \sqrt{\mu_0} H_0 \\ 0
   \end{array} \right)
 = \sqrt{\frac{\epsilon_0}{2}}\ E_0
  \psi({\bf r}_\perp;0) \left( \begin{array}{c}
   1 \\ 0 \\ 0 \\ 0 \\ 1 \\ 0 \end{array} \right) ,
 \label{eq30}
\ee
where we have made use of the~relation between amplitudes $H_0 = \sqrt{\epsilon_0/\mu_0}\ \! E_0$.
Following the~transformation relation (\ref{eq9}), and~taking into account
the functional form displayed by the~output Gaussian amplitude (\ref{eq15}), the
action of the~operator $\underline{\hat{\mathbb{U}}}$ on the~scala field $\psi$,
generates the~matrix operator

\textls[-25]{\be
 \underline{\hat{\psi}}({\bf r}_\perp;z) \equiv \hat{\underline{\mathbb{U}}}
 (\hat{\bf r}_\perp \mathbb{I} + k^{-1}\mathbb{G}_\perp, \hat{\bf p}_\perp \mathbb{I}) \psi({\bf r}_\perp;0)
   = \underline{\psi}(x\mathbb{I} + k^{-1}\mathbb{G}_x, y\mathbb{I} + k^{-1}\mathbb{G}_y; z)
  = A_z e^{\alpha_z \left( \Delta {\bf r}_\perp \mathbb{I} + k^{-1} \mathbb{G}_\perp \right)^2} ,
 \label{eq22}
\ee}
where the~argument of (\ref{eq15}) has been replaced by the~first argument of $\underline{\hat{\mathbb{U}}}$
(the second argument of $\underline{\hat{\mathbb{U}}}$ is the~generator of the~transformation from $z=0$ to
finite $z$), with $\Delta {\bf r}_\perp = {\bf r}_\perp - {\bf r}'_{\perp,0}$.
For convenience, we~have defined
\be
 \alpha_z = - \frac{1}{4\sigma_0\tilde{\sigma}_z} = \frac{ik}{2q_z} ,
\ee
with
\be
 q_z = - 2ik \sigma_0\tilde{\sigma}_z = -2ik\sigma_0\sigma_z e^{i\varphi_z}
  = 2k\sigma_0\sigma_z e^{i(\varphi_z - \pi/2)} .
 \label{qz}
\ee
From now on, because it is more compact, we shall consider the~last expression
for $q_z$ in the~derivations.

Before further proceeding, it is convenient to simplify the~argument of the~exponential in (\ref{eq22}),
just~to~determine how the~matrices involved will act on the~column vector (\ref{eq30}).
To this end, making use of the~properties satisfied by the~$\mathbb{G}$-matrices,
we find
\be
 \left( \Delta {\bf r}_\perp \mathbb{I} + k^{-1} \mathbb{G}_\perp \right)^2 =
 \Delta r_\perp^2 \mathbb{I} + 2k^{-1} \Delta {\bf r}_\perp \cdot \mathbb{G}_\perp ,
 \label{eq25}
\ee
with $\Delta r_\perp = \| \Delta {\bf r}_\perp \|$, and~where the~property (\ref{eq24}) has been used.
Now, we consider a~Taylor series expansion of the~exponential matrix in (\ref{eq22}),
where, again for convenience, the~term $\Delta {\bf r}_\perp \cdot \mathbb{G}_\perp$ is going to
be explicitly separated as $\Delta x \mathbb{G}_x + \Delta y \mathbb{G}_y$.
Thus, we have
%
\ba
 e^{\alpha_z \left( \Delta r_\perp^2 \mathbb{I} + 2k^{-1} \Delta {\bf r}_\perp \cdot \mathbb{G}_\perp \right)} & = &
 \sum_{n=0}^\infty \frac{\alpha_z^n}{n!}
  \left( \Delta r_\perp^2 \mathbb{I} + 2k^{-1} \Delta {\bf r}_\perp \cdot \mathbb{G}_\perp \right)^n
 \nonumber \\
 & = &
 \sum_{n=0}^\infty \frac{\alpha_z^n}{n!}
  \sum_{m=0}^n \left( \begin{array}{c} n \\ m \end{array} \right)
  \left( \Delta r_\perp^2 \mathbb{I} \right)^{n-m}
  \left( \frac{2}{k} \right)^m
   \left( \Delta {\bf r}_\perp \cdot \mathbb{G}_\perp \right)^m ,
 \label{eq26}
\ea
%
with
\be
 (\mu + \nu)^n = \sum_{j=0}^n \left( \begin{array}{c} n \\ j \end{array} \right) \mu^{n-j} \nu^j
 = \sum_{j=0}^n \frac{n!}{j!(n-j)!}\ \! \mu^{n-j} \nu^j .
 \label{eq27}
\ee
%
%
%

According to the~property (\ref{eq24b}), the~only surviving terms in the~second sum of (\ref{eq26})
are those with $m \le 2$, hence such an~expression can be further simplified to
\ba
 e^{\alpha_z \left( \Delta r_\perp^2 \mathbb{I} + 2k^{-1} \Delta {\bf r}_\perp \cdot \mathbb{G}_\perp \right)} & = &
 \sum_{n=0}^\infty \frac{\alpha_z^n}{n!}
  \left[ \Delta r_\perp^{2n} \mathbb{I}
  + \left( \frac{2}{k} \right) n \Delta r_\perp^{2(n-1)}
     \left( \Delta {\bf r}_\perp \cdot \mathbb{G}_\perp \right) \right.
 \nonumber \\ & & \qquad \qquad
  \left. + \left( \frac{4}{k^2} \right) n (n-1) \Delta r_\perp^{2(n-2)}
      \left( \Delta {\bf r}_\perp \cdot \mathbb{G}_\perp \right)^2 \right] ,
\ea
which can be recast in a~more compact form as
\textls[-15]{\ba
 e^{\alpha_z \left( \Delta r_\perp^2 \mathbb{I} + 2k^{-1} \Delta {\bf r}_\perp \cdot \mathbb{G}_\perp \right)}
 & = & \left( \sum_{n=0}^\infty \frac{\alpha_z^n}{n!}  \Delta r_\perp^{2n} \right)
 \left[ \mathbb{I} + \alpha_z \left( \frac{2}{k} \right) \left( \Delta {\bf r}_\perp \cdot \mathbb{G}_\perp \right)
  + \alpha_z^2 \left( \frac{4}{k^2} \right)  \left( \Delta {\bf r}_\perp \cdot \mathbb{G}_\perp \right)^2 \right]
 \nonumber \\
 & = & e^{\alpha_z \| \Delta {\bf r}_\perp \|^2} \left[ \mathbb{I} + \frac{i}{q_z} \left( \Delta {\bf r}_\perp \cdot \mathbb{G}_\perp \right)
  - \frac{1}{q_z^2} \left( \Delta {\bf r}_\perp \cdot \mathbb{G}_\perp \right)^2 \right] ,
 \label{eq29}
\ea}
%
\textls[-15]{where the~prefactor corresponds to the~Gaussian function describing the~beam (\ref{eq15}), while the~matrix terms
provide us with the~correct expression for the~electric and~magnetic fields, according to the~rule (\ref{eq9}).}

After evaluating the~combined action of the~$\mathbb{G}$-matrices and~their products on the~column
vector~(\ref{eq30}), we obtain the~expression for the~output vector field ${\bf F}$ at a~distance $z$,
\be
 {\bf F}({\bf r}_\perp;z) =
  \sqrt{\frac{\epsilon_0}{2}}\ E_0 \psi({\bf r}_\perp;z) \left( \begin{array}{c}
   1 - [(\Delta x)^2 - (\Delta y)^2]/2q_z^2 \\ \\  -\Delta x \Delta y/q_z^2 \\ \\ -\Delta x/q_z \\ \\  -\Delta x \Delta y/q_z^2 \\ \\ 1 + [(\Delta x)^2 - (\Delta y)^2]/2q_z^2 \\ \\ -\Delta y/q_z \end{array}
    \right) ,
 \label{eq32}
\ee
or, in a~more conventional fashion, in terms of the~separate electric and~magnetic components, as
%
\ba
 {\bf E}({\bf r}_\perp;z) & = &
  E_0 \psi({\bf r}_\perp;z) \left\{ \left[ 1 - \frac{(\Delta x)^2 - (\Delta y)^2}{2q_z^2} \right] {\bf \hat{x}}
      - \frac{\Delta x \Delta y}{q_z^2}\ {\bf \hat{y}}
      - \frac{\Delta x}{q_z}\ {\bf \hat{z}} \right\} ,
 \label{eq33} \\
 {\bf H}({\bf r}_\perp;z) & = &
  H_0 \psi({\bf r}_\perp;z) \left\{ - \frac{\Delta x \Delta y}{q_z^2}\ {\bf \hat{x}}
     + \left[ 1 + \frac{(\Delta x)^2 - (\Delta y)^2}{2q_z^2} \right] {\bf \hat{y}}
     - \frac{\Delta y}{q_z}\ {\bf \hat{z}} \right\} ,
 \label{eq34}
\ea
%
which include up to second-order corrections in the~transverse coordinates, with
${\bf \hat{x}}$, ${\bf \hat{y}}$ and~${\bf \hat{z}}$ denoting the~unit vectors along
the three Cartesian directions.
Next, we are going to analyze the~physical implications carried by these contributions.

Let us firstly start by neglecting the~second-order contributions.
Accordingly, the~field vector (\ref{eq32}) reads as
\be
 {\bf F}({\bf r}_\perp;z) =
  \sqrt{\frac{\epsilon_0}{2}}\ E_0 \psi({\bf r}_\perp;z) \left( \begin{array}{c}
   1 \\ 0 \\ -\Delta x/q_z \\  0 \\ 1 \\ -\Delta y/q_z \end{array}
    \right) .
 \label{eq32b}
\ee

\textls[-25]{At this level of approximation, we already find the~first-order corrections to the
paraxial approximation that lead to usual Gaussian beams, with the~electric and
magnetic fields (\ref{eq33}) and~(\ref{eq34}) being}
\ba
 {\bf E}({\bf r}_\perp;z) & = & E_0 \psi({\bf r}_\perp;z)
  \left({\bf \hat{x}} - \frac{\Delta x}{q_z}\ {\bf \hat{z}}\right)
 = E_0 \psi({\bf r}_\perp;z) \left[{\bf \hat{x}}
     + \frac{\Delta x}{2k\sigma_0 \sigma_z}\ e^{- i(\varphi_z + \pi/2)} {\bf \hat{z}}\right] ,
 \label{eq35} \\
 {\bf H}({\bf r}_\perp;z) & = & H_0 \psi({\bf r}_\perp;z)
  \left({\bf \hat{y}} - \frac{\Delta y}{q_z}\ {\bf \hat{z}}\right)
 = H_0 \psi({\bf r}_\perp;z) \left[{\bf \hat{y}}
     + \frac{\Delta y}{2k\sigma_0 \sigma_z}\ e^{- i(\varphi_z + \pi/2)} {\bf \hat{z}}\right] ,
 \label{eq36}
\ea
respectively (notice that the~$\pi/2$ phase factor is not an~extra factor, but it arises from the~expression
for $q_z$ introduced above, in Equation~(\ref{qz})).
As it can be noticed, the~two fields do not remain in-plane (i.e., contained within the~same constant
$z$ plane), as one would expect from a~usual scalar field guess, but they contain a~small out-of-plane
component, along the~longitudinal ($z$) direction.
From~these expressions, it is now possible to obtain some information about the~polarization state of the
electromagnetic field in the~near and~far fields, that is, in terms of the~value of $z$ compared to
that of~$z_R$.
Without loss of generality, in this regard, let us consider the~electric component (same holds for the
magnetic one).
The~two cases of interest worth stressing are:
\begin{itemize}[leftmargin=*,labelsep=5.8mm]
 \item[$\bullet$] If $z \ll z_R$, then $\sigma_z \approx \sigma_0$ and~$\varphi_z \approx z/z_R \ll \pi/2$.
 So, in the~near field regime, we have
\be
 {\bf E}({\bf r}_\perp;z) \approx E_0 \psi({\bf r}_\perp;z) \left({\bf \hat{x}}
     + \frac{\Delta x}{z_R}\ e^{-i\pi/2} {\bf \hat{z}}\right) ,
 \label{eq37}
\ee
and the~field is elliptically polarized, with the~polarization plane being the~$XZ$-plane
(perpendicular to the~$y$ transverse coordinate), since
\be
 \frac{E_z}{E_x} = \frac{\Delta x}{z_R}\ e^{-i\pi/2} .
\ee
As it can be noticed, only at the~center of the~beam there is horizontal polarization
(parallel~to~the~$x$ transverse coordinate); anywhere else, the~$z$-component of the
polarization increases linearly with the~distance with respect to the~center of the
beam.
Since this a~regime where the~expansion of the~beam is still negligible, this polarization
effect should be particularly relevant for distances (from the~center) of the~order of the
beam waist, $w_0$, where the~fast decrease of the~intensity might complicate its detection
(particularly, if we consider typical values used in diffraction~experiments).

\item[$\bullet$] On the~other hand, for $z \gg z_R$, in the~far field regime, we have
 $\sigma_z \approx z \sigma_0/z_R$ and~$\varphi_z \approx \pi/2$.
 Accordingly, the~electric field component reads as
\be
 {\bf E}({\bf r}_\perp;z) \approx E_0 \psi({\bf r}_\perp;z) \left({\bf \hat{x}}
     - \frac{\Delta x}{z}\ {\bf \hat{z}}\right) ,
 \label{eq38}
\ee
that is, the~field is still contained within the~$XZ$-plane, but the~elliptical polarization state
has degenerated into linear, since
\be
 \frac{E_z}{E_x} = - \frac{\Delta x}{z} ,
 \label{eq39}
\ee
which becomes negligible very quickly as $z$ increases, recovering the~linear polarization
along the~$x$ transverse coordinate at any distance from the~center of the~beam.
Therefore, contrary to the~case of a~monochromatic plane wave, transversality is not ensured
unless we consider the~limit case $\sigma_0 \to \infty$, which is precisely the~case near the
maximum of the~Gaussian, where the~wavefront could be approximated by a~nearly plane wave, as
can be seen from Equations~(\ref{eq35}) and~(\ref{eq36}), when~the~$z$-component contribution is
neglected.
This model was considered in a~previous work \cite{sanz:AnnPhysPhoton:2010} to~analyze
the problem of interference, which is revisited and~extended in next section for the~case of
interference between two Gauss-Maxwell beams.
\end{itemize}

Once we have the~correct paraxial vector field solutions to Maxwell's equations, we can
compute the~averaged paths that will be followed by photons, according to (\ref{eq12}).
In this regard, let us first start by the~lowest level of approximation, namely the~first
order in $\Delta {\bf r}_\perp$, and~compute the~time-averaged Poynting vector (\ref{avS}).
Thus, neglecting the~terms containing $(\Delta x)^2$, $(\Delta y)^2$ and~$\Delta x \Delta y$
in (\ref{eq35}) and~(\ref{eq36}), we obtain
\be
 {\bf S} = \frac{1}{2}\ E_0 H_0 |\psi|^2 \left( \begin{array}{c}
  \displaystyle \frac{\Delta x}{2k\sigma_0\sigma_z}\ \sin \varphi_z \\ \\
  \displaystyle \frac{\Delta y}{2k\sigma_0\sigma_z}\ \sin \varphi_z \\ \\
   1 \end{array} \right),
 \label{first}
\ee
where
\be
 \sin \varphi_z = \frac{z}{2k\sigma_0\sigma_z} .
\ee

The~ray equation that follows from (\ref{first}) is
\be
 \frac{d\Delta {\bf r}_\perp}{dz} = \frac{z}{4k^2\sigma_0^2}
  \frac{\Delta {\bf r}_\perp}{\sigma_z^2} ,
 \label{eq40}
\ee
where the~left-hand side term has been recast in terms of $\Delta {\bf r}_\perp$
instead of ${\bf r}_\perp$, without loss of generality (there is no effect on
the $z$-derivative).
As it can be noted, this equation corresponds to Equations~(\ref{eq18}) and~(\ref{eq19})
for the~usual Gaussian beam.
That is, we find that the~energy spreads spatially in the~same way as a~standard Gaussian
beam at the~lowest level of approximation, although the~polarization state changes
along both the~transverse and~the longitudinal directions, even if it was specifically
defined at the~input plane.

In order to determine the~deviations from the~usual Gaussian beam approach, let us
consider the~full expression for the~fields (\ref{eq35}) and~(\ref{eq36}).
The~computation of the~time-averaged Poynting vector~renders
\be
 {\bf S} = \frac{1}{2}\ E_0 H_0 |\psi|^2 \left( \begin{array}{c}
  \displaystyle \frac{\Delta x}{2k\sigma_0\sigma_z}\ \sin \varphi_z \left[ 1 + \frac{1}{2} \left( \frac{\|\Delta {\bf r}_\perp\|}{2k\sigma_0\sigma_z} \right)^2 \right] \\ \\
  \displaystyle \frac{\Delta y}{2k\sigma_0\sigma_z}\ \sin \varphi_z \left[ 1 + \frac{1}{2} \left( \frac{\|\Delta {\bf r}_\perp\|}{2k\sigma_0\sigma_z} \right)^2 \right] \\ \\
  \displaystyle 1 - \frac{1}{4} \left( \frac{\|\Delta {\bf r}_\perp\|}{2k\sigma_0\sigma_z} \right)^4 \end{array} \right),
 \label{second}
\ee
where it can be seen that the~$z$-component is diminished, precisely, in an~amount equivalent
to the~energy flux going into the~transverse directions.
Keeping in the~ray equation terms depending on the~transverse displacement, $\Delta {\bf r}_\perp$,
up to third order, we obtain
\be
 \frac{d\Delta {\bf r}_\perp}{dz} = \frac{z}{4k^2\sigma_0^2} \frac{\Delta {\bf r}_\perp}{\sigma_z^2}
  \left[ 1 + \frac{1}{2} \left( \frac{\|\Delta {\bf r}_\perp\|}{2k\sigma_0\sigma_z} \right)^2 \right] .
\ee
In order to get an~idea of the~importance of the~second term within the~square bracket,
we can evaluate it taking into account the~orders of magnitude of the~different quantities
involved in it.
To this end, we~can use values from the~experiment reported in Reference \cite{kocsis:Science:2011}.
Accordingly, in meters, $\lambda \sim 10^{-6}$, while $\sigma_0$, $\sigma_z$ and
$\|\Delta {\bf r}_\perp\|$ are the~three of the~order of $\sim 10^{-3}$.
With these values, the~correction term is of the~order of $10^{-6}$,
which fully justifies that, in a~good approximation, a~scalar Gaussian
beam description can be used instead of a~more exact vector approach.


\subsection{Arbitrarily Polarized Gauss-Maxwell Beams}
\label{sec33}

In the~previous section, for simplicity, we have considered a~simple case of linear
polarization at the~input plane.
As seen, the~polarization of this beam changes with $z$.
Now we are going to extend the~approach to a~general input polarization state with
the polarization plane perpendicular to the~longitudinal direction.
Such a~state can be described by the~superposition
\be
 |P\rangle = \cos (\theta/2) |H\rangle + \sin (\theta/2) e^{i\phi} |V\rangle ,
 \label{eq69}
\ee
with $\theta \in (0,\pi)$ and~$\phi \in (0, 2\pi)$ defined on the~Poincar\'e sphere
\cite{collett-spie}, and~where the~vector states $|H\rangle$ ($|P\rangle$ for $\theta = 0$)
and $|V\rangle$ ($|P\rangle$ for $\theta = \pi$) denote, respectively, horizontal and~vertical
polarization with respect to the~$x$-axis.
For simplicity, from now on, we shall consider the~coefficients $\alpha = \cos (\theta/2) e^{-i\delta/2}$
and $\beta = \sin (\theta/2) e^{i\phi + i\delta/2}$, which also include any possible relative phase shift
imprinted on the~state by some external action (e.g., as a~result of a~weak measurement).
As it can be noticed, these coefficients satisfy the~relation $|\alpha|^2 + |\beta|^2 = 1$ (unit radius
on the~Poincar\'e sphere).

To start with, let us consider the~input electric and~magnetic field vectors
\ba
 {\bf E}({\bf r}_\perp;0) & = & E_0 \psi({\bf r}_\perp;0) \left( \begin{array}{c}
    \alpha \\ \beta \\ 0 \end{array} \right) ,
 \nonumber \\
 {\bf H}({\bf r}_\perp;0) & = & H_0 \psi({\bf r}_\perp;0) \left( \begin{array}{c}
    -\beta \\ \alpha \\ 0 \end{array} \right) ,
 \label{eq59}
\ea
which are expressed in terms of an~arbitrary polarization vector field specified by the
coefficients $\alpha$ and~$\beta$ introduced above.
The~corresponding six-dimensional electromagnetic vector field at the~input plane $z=0$ is
\be
 {\bf F}({\bf r}_\perp;0) = \sqrt{\frac{\epsilon_0}{2}}\ E_0
  \psi({\bf r}_\perp;0) \left( \begin{array}{c}
   \alpha \\ \beta \\ 0 \\ -\beta \\ \alpha \\ 0 \end{array} \right) ,
 \label{eq60}
\ee
which reduces to (\ref{eq30}) when $\alpha = 1$.
Taking into account how the~different $\mathbb{G}$-matrices and~their products operate
over the~column vector (\ref{eq60}),
the expression for the~propagated polarized vector field becomes
\be
 {\bf F}({\bf r}_\perp;z) =
  \sqrt{\frac{\epsilon_0}{2}}\ E_0 \psi({\bf r}_\perp;z) \left( \begin{array}{c}
   \alpha - \alpha [(\Delta x)^2 - (\Delta y)^2]/2q_z^2 - \beta \Delta x \Delta y/q_z^2 \\ \\
   \beta  + \beta [(\Delta x)^2 - (\Delta y)^2]/2q_z^2  - \alpha \Delta x \Delta y/q_z^2 \\ \\
  -(\alpha \Delta x + \beta \Delta y)/q_z \\ \\
  -\beta  + \beta [(\Delta x)^2 - (\Delta y)^2]/2q_z^2  - \alpha \Delta x \Delta y/q_z^2 \\ \\
   \alpha + \alpha [(\Delta x)^2 - (\Delta y)^2]/2q_z^2 + \beta \Delta x \Delta y/q_z^2 \\ \\
   (\beta \Delta x - \alpha \Delta y)/q_z
  \end{array} \right) .
 \label{eq62}
\ee

As before, from this vector we readily obtain the~electric and~magnetic vector fields along
the~$z$-axis:
%
\ba
 {\bf E}({\bf r}_\perp;z) & = & E_0 \psi({\bf r}_\perp;z) \left\{
   \left( \alpha - \frac{\alpha [(\Delta x)^2 - (\Delta y)^2]}{2q_z^2} - \frac{\beta \Delta x \Delta y}{q_z^2} \right) {\bf \hat{x}}
 \right. \nonumber \\ & & \left. \qquad \quad
 + \left( \beta  + \frac{\beta [(\Delta x)^2 - (\Delta y)^2]}{2q_z^2}  - \frac{\alpha \Delta x \Delta y}{q_z^2} \right) {\bf \hat{y}}
 - \frac{\alpha \Delta x + \beta \Delta y}{q_z}\ {\bf \hat{z}} \right\} ,
 \label{eq63} \\
 {\bf H}({\bf r}_\perp;z) & = & H_0 \psi({\bf r}_\perp;z) \left\{
 - \left[ \beta  - \frac{\beta [(\Delta x)^2 - (\Delta y)^2]}{2q_z^2}  + \frac{\alpha \Delta x \Delta y}{q_z^2} \right] {\bf \hat{x}}
 \right. \nonumber \\ & & \left. \qquad \quad
 + \left[ \alpha + \frac{\alpha [(\Delta x)^2 - (\Delta y)^2]}{2q_z^2} + \frac{\beta \Delta x \Delta y}{q_z^2} \right] {\bf \hat{y}}
 + \frac{\beta \Delta x - \alpha \Delta y}{q_z}\ {\bf \hat{z}} \right\} .
 \label{eq64}
\ea

As seen above, dealing with these full expressions does not lead to any important
difference with respect to retaining only up to the~first order in the~transverse
displacement, $\Delta {\bf r}_\perp$, and, in the~present case, may lead to a~rather
complex expression for the~time-averaged Poynting vector (which does not imply any significant
advantage both at the~conceptual level and~at the~methodological one).
Therefore, let us thus keep terms at the~lowest level in $\Delta {\bf r}_\perp$,
which reduces the~six-component electromagnetic vector field to
\be
 {\bf F}({\bf r}_\perp;z) =
  \sqrt{\frac{\epsilon_0}{2}}\ E_0 \psi({\bf r}_\perp;z) \left( \begin{array}{c}
   \alpha \\
   \beta  \\
  -(\alpha \Delta x + \beta \Delta y)/q_z \\
  -\beta  \\
   \alpha \\
   (\beta \Delta x - \alpha \Delta y)/q_z
  \end{array} \right) ,
 \label{eq62b}
\ee
and the~electric and~magnetic fields to
\ba
 {\bf E}({\bf r}_\perp;z) & = & E_0 \psi({\bf r}_\perp;z) \left[
   \alpha {\bf \hat{x}} + \beta {\bf \hat{y}} + \frac{\alpha \Delta x + \beta \Delta y}{2k\sigma_0\sigma_z}\ e^{-i(\varphi_z + \pi/2)} {\bf \hat{z}} \right] ,
 \label{eq63b} \\
 {\bf H}({\bf r}_\perp;z) & = & H_0 \psi({\bf r}_\perp;z) \left[
 - \beta {\bf \hat{x}} + \alpha {\bf \hat{y}} + \frac{\beta \Delta x - \alpha \Delta y}{2k\sigma_0\sigma_z}\ e^{-i(\varphi_z - \pi/2)} {\bf \hat{z}} \right] .
 \label{eq64b}
\ea

In order to determine the~flux of associated rays, we first compute the~time-averaged
Poynting vector from the~fields (\ref{eq63b}) and~(\ref{eq64b}), which reads as
\ba
 {\bf S} & = & \frac{1}{2}\ E_0 H_0 |\psi|^2 \left(
   \begin{array}{c}
    \displaystyle
     \frac{\Delta x}{2k\sigma_0\sigma_z}\ \sin \varphi_z
     + {\rm Im} (\alpha \beta^*)\ \frac{\Delta y}{k\sigma_0\sigma_z}\ \cos \varphi_z \\ \\
    \displaystyle
     \frac{z \Delta y}{2k\sigma_0\sigma_z}\ \sin \varphi_z
     - {\rm Im} (\alpha \beta^*)\ \frac{\Delta x}{k\sigma_0\sigma_z^2}\ \cos \varphi_z \\ \\
    1 \end{array} \right)
 \nonumber \\
 & = & \frac{1}{2}\ E_0 H_0 |\psi|^2 \left(
   \begin{array}{c}
    \displaystyle
     \frac{z \Delta x}{4k^2\sigma_0^2\sigma_z^2}
     - \sin \theta \sin (\phi + \delta)\ \frac{z_R \Delta y}{4k^2\sigma_0^2\sigma_z^2} \\ \\
    \displaystyle
     \frac{z \Delta y}{4k^2\sigma_0^2\sigma_z^2}
     + \sin \theta \sin (\phi + \delta)\ \frac{z_R \Delta x}{4k^2\sigma_0^2\sigma_z^2} \\ \\
    1 \end{array} \right) ,
 \label{eq70}
\ea
with
\be
 \cos \varphi_z = \frac{z_R}{2k\sigma_0\sigma_z} .
\ee

From (\ref{eq70}), we obtain the~following set of ray equations
\ba
 \frac{d\Delta x}{dz} & = & \frac{z \Delta x}{4k^2\sigma_0^2\sigma_z^2}
  - \sin \theta \sin (\phi + \delta)\ \frac{z_R \Delta y}{4k^2\sigma_0^2\sigma_z^2} ,
 \label{eqraypol1} \\
 \frac{d\Delta y}{dz} & = & \frac{z \Delta y}{4k^2\sigma_0^2\sigma_z^2}
  + \sin \theta \sin (\phi + \delta)\ \frac{z_R \Delta x}{4k^2\sigma_0^2\sigma_z^2} ,
 \label{eqraypol2}
\ea
which are coupled by virtue of the~polarization factor $\sin \theta \sin (\phi + \delta)$.
As it can readily be seen, if the~polarization state of the~input beam is linear, regardless
of its vibration direction \mbox{($\phi = 0$ or $\pi$),} and~has not received any extra kick ($\delta = 0$),
then the~coupling disappears and~we recover the~same situation that we had in the~previous section
for the~horizontally polarized input Gaussian beam.
Otherwise, if the~polarization state is not either horizontal or vertical, the~second term in the
polarization factor provides an~extra spin component that makes the~rays to get out the
plane where they are initially contained, namely the~$XZ$-plane or the~$YZ$-plane.
Accordingly, we have a~clear description of the~energy flux without appealing to a~standard
picture of rotating electric field vectors, where the~electromagnetic energy streamlines
already display such a~rotation following the~spin imprinted by the~polarization vector.
For instance, in the~case of circularly right-handed polarized light ($\theta = \pi/2$ and
$\phi = -\pi/2$, with $\delta = 0$), described by the~(non-normalized) polarization vector
\be
 |P\rangle = |H\rangle - i |V\rangle ,
 \label{eq69bb}
\ee
which is the~case for which the~polarization factor is maximal, we have
\ba
 \frac{d\Delta x}{dz} & = & \frac{z \Delta x}{4k^2\sigma_0^2\sigma_z^2}
  + \frac{z_R \Delta y}{4k^2\sigma_0^2\sigma_z^2} ,
 \label{eqraypol1bb} \\
 \frac{d\Delta y}{dz} & = & \frac{z \Delta y}{4k^2\sigma_0^2\sigma_z^2}
  - \frac{z_R \Delta x}{4k^2\sigma_0^2\sigma_z^2} ,
 \label{eqraypol2bb}
\ea
which can be recast as
\be
 \frac{d\Delta {\bf r}_\perp}{dz} = \mathbb{R}_{\varphi_z} \Delta {\bf r}_\perp ,
\ee
with $\mathbb{R}_{\varphi_z}$ being the~rotation matrix
\be
 \mathbb{R}_{\varphi_z} = \left( \begin{array}{cc}
    \sin \varphi_z & \cos \varphi_z \\
    - \cos \varphi_z & \sin \varphi_z
  \end{array} \right)
  = \left( \begin{array}{cc}
    \cos (\varphi_z - \pi/2) & - \sin (\varphi_z - \pi/2) \\
    \sin (\varphi_z - \pi/2) & \cos (\varphi_z - \pi/2),
  \end{array} \right)
\ee
that is, the~position vector $\Delta {\rm r}_\perp$ is acted by a~rotation
$\varphi_z - \pi/2$.


\section{Young-Type Interference with Gauss-Maxwell Beams}
\label{sec4}



Now we are going to consider the~case of a~coherent superposition of two
linearly polarized Gaussian beams at the~input plane $z=0$.
This is the~case, for instance, of the~two diffracted beams produced by
two slits (Young-type diffraction), which arise from the~incidence of light
of a~conventional laser pointer linearly polarized.
As mentioned above, the~standard Gaussian beam description can be used to
describe the~eventual spatial distribution of radiation, but it does not
provide us with any clue on polarization, in case we need it.
That is, we would have a~two-beam coherent superposition
\be
 \psi({\bf r}_\perp;z) = \psi_1 ({\bf r}_\perp;z) + \psi_2 ({\bf r}_\perp;z) ,
 \label{eq42}
\ee
with
\be
 \psi_i({\bf r}_\perp;z) =
   \mathcal{N} \left( \frac{1}{2\pi\tilde{\sigma}_{z,i}^2} \right)^{1/2}
   e^{-\|{\bf r}_\perp - {\bf r}_{\perp,c}^{(i)} \|^2/4\sigma_{0,i}\tilde{\sigma}_{z,i}} ,
 \label{eq41}
\ee
denoting the~diffracted Gaussian beams, with $i=1,2$.
For simplicity, although without loss of generality, we are going to consider that
both beams are identical ($\sigma_{0,i} = \sigma_0$), they are symmetrically placed
with respect to $x=0$ ($x_{0,1} = - x_{0,2} = x_c$), and~aligned along the~$x$-axis
\mbox{($y_{0,1} = y_{0,2} = 0$).}
As~for~the~prefactor $\mathcal{N}$ in (\ref{eq41}), it is just a~$z$-independent
global norm factor that amounts to approximately $1/\sqrt{2}$ when the~overlap
of the~two input beams is nearly zero, as it can readily be seen from its functional
form,
\be
 \mathcal{N} = \frac{1}{\sqrt{2}} \frac{1}{\sqrt{1 + e^{-x_c^2/2\sigma_0^2}}} .
 \label{eqB10}
\ee

Without any need for considering the~effect of polarization (which is the~same to
say that both beams are equally polarized, as we have already seen above), the
intensity of the~radiation described by the~scalar superposition field (\ref{eq42}),
except for constant factors, is going to be proportional to the~density profile along
the $x$-direction,
\be
 |\psi|^2 = |\psi_1|^2 + |\psi_2|^2 + 2 {\rm Re} \left( \psi_1 \psi_2^* \right)
  \sim e^{- (x - x_c)^2/2\sigma_z^2} + e^{- (x + x_c)^2/2\sigma_z^2} +
  2 e^{- (x^2 + x_c^2)/2\sigma_z^2} \cos \left( \kappa_z x \right) ,
 \label{density1}
\ee
where
\be
 \kappa_z = \frac{x_c z}{\sigma_z^2 z_R} .
\ee

Note that we have denoted (\ref{density1}) with the~term density, because of its
direct relation to the~time-averaged electromagnetic energy density (except for a
proportionality constant) \cite{bornwolf-bk}.
Furthermore, it can readily be seen that, in the~limit $z \gg z_R$, where
large $x$ values are involved ($x \gg x_c$), the~density (\ref{density1}) can be
approximated to
\be
 |\psi|^2 \sim 4 e^{- z_R^2 x^2/2\sigma_0^2 z^2}
  \cos^2 \left( \frac{2\pi x_c x}{\lambda z} \right) ,
 \label{density2}
\ee
with
\be
 \kappa_z \approx \frac{x_c z_R}{\sigma_0^2 z} = \frac{4\pi x_c}{\lambda z} .
\ee

The~density (\ref{density2}) describes a~Young-type fringe pattern (along the~$x$-direction)
modulated by a~Gaussian envelop, arising from the~Gauss shape of the~input beams, and~with
maxima at positions
\be
 x_n = \frac{\lambda z}{d}\ n ,
 \label{quant}
\ee
where $d = 2 x_c$.
Actually, if we assume that $x_n/z = \sin \theta_n$, with $\theta_n$ being the~angular
position of each maximum with respect to the~coordinate origin at the~input plane, the
above condition becomes
\be
 d \sin \theta_n = \lambda n ,
\ee
that is, the~usual two-slit interference condition \cite{bornwolf-bk}.

Let us now consider the~case of polarized input beams, particularly, the
case of horizontal polarization (in next section, we shall introduce the
case of arbitrarily polarized superimposed beams).
Following the~procedure described in Section~\ref{sec32}, we find
that the~electric and~magnetic fields (within the~paraxial
approximation) associated with each partial wave (\ref{eq41}) are
\ba
 {\bf E}_i({\bf r}_\perp;z) & = & E_0 \psi_i({\bf r}_\perp;z)
   \left[{\bf \hat{x}} + \frac{\Delta x_i}{2k\sigma_0\sigma_z}\ e^{-i(\varphi_z + \pi/2)} {\bf \hat{z}}\right] ,
 \label{eq43} \\
 {\bf H}_i({\bf r}_\perp;z) & = & H_0 \psi_i({\bf r}_\perp;z)
   \left[{\bf \hat{y}} + \frac{\Delta y_i}{2k\sigma_0\sigma_z}\ e^{-i(\varphi_z + \pi/2)} {\bf \hat{z}}\right] ,
 \label{eq44}
\ea
where, again, we have assumed identical beams ($E_{0,1} = E_{0,2} = E_0$ and
$H_{0,1} = H_{0,2} = H_0$).
The~total electric and~magnetic fields that result from the~superposition of
both beams read as
\ba
 {\bf E}({\bf r}_\perp;z) & = & E_0 \left\{ \psi({\bf r}_\perp;z) {\bf \hat{x}}
  + \frac{e^{-i(\varphi_z + \pi/2)}}{2k\sigma_0\sigma_z}
  \left[ x \psi({\bf r}_\perp;z) - x_c \Delta \psi({\bf r}_\perp;z) \right] {\bf \hat{z}} \right\}
 \label{eq45} \\
 {\bf H}({\bf r}_\perp;z) & = & H_0 \left\{ \psi({\bf r}_\perp;z) {\bf \hat{y}}
  + \frac{e^{-i(\varphi_z + \pi/2)}}{2k\sigma_0\sigma_z}\ y \psi({\bf r}_\perp;z) {\bf \hat{z}} \right\} ,
 \label{eq46}
\ea
where $\Delta x_i = x - x_{c,i}$ and~$\Delta \psi = \psi_1 - \psi_2$.
From these expressions, we now compute the~time-averaged Poynting vector, which reads as
\begin{small}
\textls[-25]{\ba
 {\bf S} & = & \frac{1}{2}\ E_0 H_0 \left(
   \begin{array}{c}
    \displaystyle
     \frac{\sin \varphi_z}{2k\sigma_0\sigma_z} \left[ x |\psi|^2 - x_c \left( |\psi_1|^2 - |\psi_2|^2 \right) \right]
     + \frac{\cos \varphi_z}{k\sigma_0\sigma_z}\ x_c {\rm Im} \left( \psi_1 \psi_2^* \right) \\ \\
    \displaystyle
     \frac{\sin \varphi_z}{2k\sigma_0\sigma_z}\ y |\psi|^2 \\ \\
    |\psi|^2 \end{array} \right)
  \\
 & = & \frac{1}{2}\ E_0 H_0\left(
   \begin{array}{c}
    \displaystyle
     \frac{z}{4k^2\sigma_0^2\sigma_z^2} \left[ \Delta x_1 |\psi_1|^2 + \Delta x_2 |\psi_2|^2
     + 2 x {\rm Re} \left( \psi_1 \psi_2^* \right) \right]
     + \frac{z_R}{4k^2\sigma_0^2\sigma_z^2}\ x_c {\rm Im} \left( \psi_1 \psi_2^* \right) \\ \\
    \displaystyle
     \frac{z}{4k^2\sigma_0^2\sigma_z^2}\ y |\psi|^2 \\ \\
    |\psi|^2 \end{array} \right) , \label{eq49}
 \nonumber \\
 \nonumber
\ea}
\end{small}

As before, the~ray equation is now obtained from the~components of (\ref{eq49}),
since flow of the~beam at each point is described by the~direction of the~time-averaged
Poynting vector right on that point, which is also the~direction along which the~wave
vector ${\bf k}$ points \cite{bornwolf-bk,barnett:PhilTRSA:2010}.
Accordingly, we obtain
\ba
 \frac{dx}{dz} & = &
    \frac{z}{4k^2\sigma_0^2\sigma_z^2} \left[ \Delta x_1\ \frac{|\psi_1|^2}{|\psi|^2}
    + \Delta x_2\ \frac{|\psi_2|^2}{|\psi|^2}
    + 2 x \frac{{\rm Re} \left( \psi_1 \psi_2^* \right)}{|\psi|^2} \right]
    + \frac{z_R x_c}{2k^2\sigma_0^2\sigma_z^2}\ \frac{{\rm Im} \left( \psi_1 \psi_2^* \right)}{|\psi|^2}
 \nonumber \\
 & = & \frac{z}{4k^2\sigma_0^2\sigma_z^2} \left[ \frac{
  \Delta x_1\ e^{- (\Delta x_1)^2/2\sigma_z^2} + \Delta x_2\ e^{- (\Delta x_2)^2/2\sigma_z^2}
    + 2 x e^{- (x^2 + x_c^2)/2\sigma_z^2} \cos \left( \kappa_z x \right)}
  {e^{- (\Delta x_1)^2/2\sigma_z^2} + e^{- (\Delta x_2)^2/2\sigma_z^2} + 2 e^{- (x^2 + x_c^2)/2\sigma_z^2} \cos \left( \kappa_z x \right)} \right]
 \nonumber \\
 & & - \frac{z_R x_c}{2k^2\sigma_0^2\sigma_z^2} \left[
    \frac{e^{- (x^2 + x_c^2)/2\sigma_z^2} \sin \left( \kappa_z x \right)}
  {e^{- (\Delta x_1)^2/2\sigma_z^2} + e^{- (\Delta x_2)^2/2\sigma_z^2} + 2 e^{- (x^2 + x_c^2)/2\sigma_z^2} \cos \left( \kappa_z x \right)} \right] ,
 \label{eq54} \\
 \frac{dx}{dz} & = & \frac{z}{4k^2\sigma_0^2\sigma_z^2}\ y ,
 \label{eq55}
\ea
%
where the~last two terms in (\ref{eq54}) depend on the~effective spatial frequency
$\kappa_z$, as (\ref{density1}).
It can be noticed from Equation~(\ref{eq54}) that when the~wave function $\psi$ is formed by
only one beam, it reduces to the~same expression we had for a~single Gaussian beam.
The~presence of both beams, however, gives rise to a~rather complex evolution, as
described in Reference \cite{sanz:JPA:2008}, which is highly nonlinear \cite{luis:AOP:2015} and
does not allow for analytical solutions.
This is in contrast with the~evolution of the~rays along the~$y$-direction, which is fully
analytical and~corresponds to the~result that we have already found in Section~\ref{sec22}
for unpolarized standard Gaussian beams and~in Section~\ref{sec32} for linearly polarized
Gauss-Maxwell beams.
Yet in the~far field it is possible to perform some guesses of interest from Equation~(\ref{eq54})
without even solving it.
Thus, if we assume $z \gg z_R$ (and $x \gg x_c$), this equation reads as
\be
 \frac{dx}{dz} \approx \frac{x}{z} - \frac{z_R x_c}{2z^2} \left[
    \frac{\sin \left( \kappa_z x \right)}{1 + \cos \left( \kappa_z x \right)} \right] .
 \label{eq54app}
\ee

According to this equation, the~transverse momentum along the~$x$-direction increases linearly
with $x$ at any sufficiently distant output plane $z$ (with respect to the~input plane), from
negative values to positive ones.
This behavior is governed by the~first term in the~equation (the second term decays as $z^{-2}$), which
is interrupted whenever the~denominator of the~second term cancels out, that is, $\kappa_z x = (2n+1)\pi$.
In this case, the~second term becomes the~leading one, which gives rise to some sudden variations of
the transverse momentum (a sort of spiky behavior) at those points, positive for $x<0$ and
negative for $x>0$.
Regarding the~rays that can be expected, if we neglect this second term, the~equation describes
straight lines; taking into account the~second term, on the~other hand, the~rays will be grouped
in swarms, each one lying within two consecutive ``spikes.''
Although, as said above, this is just a~guess, note that it coincides with both previous models
\cite{sanz:AnnPhysPhoton:2010,dimic:PhysScr:2013} and~the experiment reported in Reference \cite{kocsis:Science:2011}.



Now, the~interference with two polarized beams can be extended to a~general case where the
polarization state associated with each Gaussian beam is arbitrary.
This involves an~interesting subtlety worth emphasizing:
different polarizations give rise to a~loss of mutual coherence, related to both the~so-called
Arago-Fresnel laws of interference with polarized light in optics \cite{sanz:JRLR:2010} and, within
the quantum realm, the~well-known problem of the~quantum (which-way) erasure \cite{walborn:PRA:2002}.
To better appreciate this fact, notice that the~electric and~magnetic components for each field are
now
\ba
 {\bf E}_i({\bf r}_\perp;z) & = & E_0 \psi_i({\bf r}_\perp;z)
   \left[ \alpha_i {\bf \hat{x}} + \beta_i {\bf \hat{y}}
   + \frac{\left( \alpha_i \Delta x_i + \beta_i \Delta y_i \right)}{2k\sigma_0\sigma_z}\ e^{-i(\varphi_z + \pi/2)} {\bf \hat{z}}\right] ,
 \label{eq65} \\
 {\bf H}_i({\bf r}_\perp;z) & = & H_0 \psi_i({\bf r}_\perp;z)
   \left[ - \beta_i {\bf \hat{x}} + \alpha_i {\bf \hat{y}}
   + \frac{\left( \beta_i \Delta x_i - \alpha_i \Delta y_i \right)}{2k\sigma_0\sigma_z}\ e^{-i(\varphi_z + \pi/2)} {\bf \hat{z}}\right] ,
 \label{eq66}
\ea
where we can appreciate that coordinates and~polarization coefficients are both intertwined,
which does not allow us to recast the~electromagnetic field as a~factorizable function of
coordinates and~polarization states, that is, in general terms, something of the~kind $|\psi_i\rangle |P_i\rangle$,
which would simplify the~final solution.
From the~corresponding total electric and~magnetic fields arising from the~superposition (their~expressions are
provided in Appendix~\ref{appB}), it is shown that the~intensity distribution at a~distance $z$ from the~input plane is
proportional to the~density
\be
 |\psi|^2 = |\psi_1|^2 + |\psi_2|^2 + 2 {\rm Re} \left[ \langle P_2 | P_1 \rangle \psi_1 \psi_2^* \right] ,
 \label{density1b}
\ee
which involves the~factor
\ba
 \langle P_2 | P_1 \rangle & = & \alpha_1 \alpha_2^* + \beta_1 \beta_2^*
 \nonumber \\
  & = & \cos \theta_1/2 \cos \theta_2/2 e^{-i(\delta_1 - \delta_2)/2}
      + \sin \theta_1/2 \sin \theta_2/2 e^{i(\phi_1 - \phi_2) + i (\delta_1 - \delta_2)/2} .
 \label{overlap}
\ea

This factor  leads to a~reduction of the~fringe visibility
as the~polarization states $|P_1\rangle$ and~$|P_2\rangle$ become more different, with the~maximum loss,
with $\langle P_2 | P_1 \rangle$, when these states are orthogonal (e.g.,~horizontal vs. vertical polarization,
or left-handed vs. right-handed circular polarization), in~agreement with the~aforementioned Arago-Fresnel
laws.

\textls[-5]{By inspecting (\ref{density1b}), we readily notice that, again, the~fringe distribution is along the~$x$ coordinate.
In the~case of the~horizontally polarized beams discussed above, we could get an~idea of why this was the
behavior: leaving aside the~scalar fields themselves, the~total electric and~magnetic fields explicitly depended
on only one transverse coordinate, namely the~electric field on the~$x$-coordinate and~the magnetic field on the
$y$-coordinate, as it can be seen in Equations~(\ref{eq45}) and~(\ref{eq46}), respectively.
This~is~in clear contrast with the~fields described by Equations~(\ref{eq65b}) and~(\ref{eq66b}), which both depend on
both coordinates (in their $z$-component).
That is, not only coordinates and~polarization coefficients are mixed up, but there is also a~coupling
between different field components, which is going to play an~interesting role from the~point of view of the
topology displayed by the~corresponding rays.
Such a~mixture of coordinates becomes more apparent from the~time-averaged Poynting vector or the~corresponding
ray equations along the~$x$ and~$y$ directions in terms of the~longitudinal coordinate $z$ \mbox{(see expressions in
Appendix~\ref{appB}).}
The~latter equations reduce to Equations~(\ref{eq54}) and~(\ref{eq55}), respectively, in the~particular case
$\alpha_1 = \alpha_2 = 1$ and~$\beta_1 = \beta_2 = 0$ (i.e., two input beams horizontally
polarized at the~input plane $z=0$).
However, unlike the~case described by Equations~(\ref{eq54}) and~(\ref{eq55}), now~the~generality of the~current
equations makes difficult to get a~clear clue on what is going on with the~dynamics they describe.
Nonetheless, still we notice that there is a~correlation between transverse coordinates (mediated,
in turn, by the~coupling with the~polarization state) that may induce out-of-plane dynamics (i.e.,
dependence of one transverse component of the~rays on the~other, and~vice~versa), which~are not
present in the~case of either horizontal or vertical polarized input beams \cite{sanz:AnnPhysPhoton:2010}.}



\section{Final Remarks}
\label{sec5}

In this work we have explored a~reliable methodology to tackle the~study and
analysis of experiments performed with Gaussian beams from a~ray-based
perspective.
We could have just remain at the~level of standard Gaussian beams, which are acceptable
paraxial solutions to Helmholt'z equation and, as it has been shown, they admit a~simple
treatment in terms of such rays.
However, despite its convenience, this approach is limited by the~fact that it does not
include the~description of the~polarization state of light, which requires a~vector
treatment.
With such a~purpose, we have considered the~theoretical framework provided by the~approach
developed by Mukunda {et al}.\ \cite{mukunda:JOSAA:1986} of Gauss-Maxwell beams, which
are paraxial vector solutions to Maxwell's equations.
Accordingly, we have a~formulation that nicely combines the~field distribution in coordinates
at the~same time that accounts for the~polarization state, optimal whenever polarization plays
a major role, as it is the~case, for instance, in the~experiment on weak measurements reported
by Kocsis {et al.}\ \cite{kocsis:Science:2011}.

Here, in particular, because light undergoes fast oscillations with time, we have considered
a time-averaged approach, which has allowed us to determine rays analogous to those provided
by geometrical optics.
However, contrary to geometrical rays, the~advantage of the~rays here introduced is that they
follow the~evolution of the~electromagnetic field and, therefore, provide us with an~accurate
description of diffraction and~interference phenomena.
This is achieved by relating them with the~components of the~time-averaged Poynting vector,
which describe how electromagnetic energy distribute spatially.
Of course, because we have considered paraxial conditions, we have been able to relate this
distribution with the~longitudinal component of such a~vector instead of assuming a~full
time-dependent picture --- which would be more accurate as well as more complicated, without,
however, adding any new physics.
Nonetheless, it is worth mentioning that the~ray-based picture here provided is able to account
for results like those reported in Reference \cite{kocsis:Science:2011} without further appealing to quantum
mechanical notions, since the~basic ingredients are already contained in standard (classical)
electromagnetism.
In other words, although Maxwell's equations say nothing about probabilities, still~they are
useful to reproduce behaviors typically associated with quantum mechanics.
Actually, without getting too deeper into the~question of whether photons have or have not a
wave function, we note that the~above approach is somehow in compliance with Bohm's view (see
p.~98 in Reference \cite{bohm-bk:QTh}) that, in absence of currents, like the~quantum probability density,
electromagnetic energy (light) ``acts~like a~fluid, which flows continuously without loss or gain
from one point to another''.
Accordingly, the~rays here defined just play the~role of the~corresponding streamlines that allow
us to understand how the~energy spatially transfers from one place to another in presence of
diffraction, interference and~polarization, the~three typical traits of electromagnetic optics.


\vspace{12pt}

\authorcontributions{All authors have read and agree to the published version of the manuscript.
Conceptualization, A.S., M.D. and M.B.; methodology, A.S., M.D. and M.B.; project administration, A.S.;
validation, A.S., M.D. and M.B.; formal analysis, A.S., M.D. and M.B.; investigation, A.S., M.D. and M.B.;
writing--original draft preparation, A.S.; writing--review and editing, A.S., M.D. and M.B.; validation, A.S.;
visualization, A.S., M.D. and M.B.; supervision, A.S.; funding acquisition, A.S., M.D. and M.B.


\vspace{5pt}
\noindent
{\bf Funding:} This research and the APC have been both funded by the Spanish Agencia Estatal de Investigaci\'on (AEI)
and the European Regional Development Fund (ERDF) grant number FIS2016-76110-P.}


\conflictsofinterest{Authors declare no conflict of interest.}


\appendixtitles{yes}
\appendix

\section{General Solution to the~Paraxial Equation (\ref{eq4})}
\label{appA}

\textls[-5]{The~integral (\ref{eq14}) is the~general solution to the~paraxial Equation (\ref{eq4}),
as it was mentioned in Section~\ref{sec3}.
To prove this, let us consider without loss of generality the~one-dimensional case,
and then we shall proceed with the~generalization to two dimensions.
In such a~case, Equation~(\ref{eq4}) can be recast as}
\be
 i\ \frac{\partial \psi(x;z)}{\partial z} = - \frac{1}{2k}
  \frac{\partial^2 \psi (x;z)}{\partial x^2} .
 \label{eqA1}
\ee

The~solution to $\psi(x;z)$ can be expressed as a~linear combination of plane waves,
as
\be
 \psi(x;z) = \frac{1}{\sqrt{2\pi}} \int e^{i\kappa x}\ \! \tilde{\psi}(\kappa;z) d\kappa .
 \label{eqA2}
\ee

If this ansatz is substituted into Equation~(\ref{eqA1}), then we obtain the~following
simpler equation:
\be
 i\ \frac{\partial \tilde{\psi}(\kappa;z)}{\partial z} = - \frac{\kappa^2}{2k}\ \! \tilde{\psi}(\kappa;z) ,
 \label{eqA3}
\ee
which solution
\be
 \tilde{\psi}(\kappa;z) = e^{-i\kappa^2 z/2k}\ \! \tilde{\psi}(\kappa;0) .
 \label{eqA4}
\ee

Now, this latter expression is substituted into the~integrand of the~ansatz (\ref{eqA2}), rendering
\be
 \psi(x;z) = \frac{1}{\sqrt{2\pi}} \int e^{i\kappa x - i\kappa^2 z/2k}\ \! \tilde{\psi}(\kappa; 0) d\kappa .
 \label{eqA5}
\ee

Since the~amplitude $\psi(x;0)$ at $z=0$ is assumed to be known, $\tilde{\psi}(\kappa;0)$ will also be
known according to the~inverse of the~transformation (\ref{eqA2}) for $z=0$:
\be
 \tilde{\psi}(\kappa;0) = \frac{1}{\sqrt{2\pi}} \int e^{-i\kappa x}\ \! \psi(x;0) dx .
 \label{eqA6}
\ee

Substituting this expression into (\ref{eqA5}) renders
\be
 \psi(x;z) = \frac{1}{2\pi} \int e^{i\kappa (x - x') - i\kappa^2 z/2k}\ \! \psi(x';0) d\kappa dx' .
 \label{eqA5b}
\ee

The~integral over $\kappa$ can be easily done to yield
\be
 \psi(x;z) = \frac{e^{-i\pi/4}}{\sqrt{\lambda z}} \int e^{ik(x - x')/2z}\ \! \psi(x';0) dx' ,
 \label{eqA7}
\ee
\textls[-45]{which accounts for the~transverse propagation of the~amplitude $\psi$ as a~function of the
(longitudinal) $z$ coordinate.}

The~generalization of the~integral (\ref{eqA7}) to two dimensions (the case considered in
Section~\ref{sec3}) is~straightforward.
Given there is no correlation between the~$x$ and~$y$ coordinates, the~full solution is just
the direct product of (\ref{eqA7}) evaluated for $x$ and~$y$, respectively:
\be
 \psi({\bf r}_\perp;z) = \frac{1}{i\lambda z}  \int e^{ik\|{\bf r}_\perp - {\bf r}'_\perp \|^2/2z}\ \! \psi({\bf r}'_\perp;0) d{\bf r}'_\perp .
 \label{eqA8}
\ee
As for the~full solution, $\Psi({\bf r}_\perp;z)$, it can be readily obtained for the~latter by
including the~corresponding translational exponential factor:
\be
 \Psi({\bf r}) = \psi({\bf r}_\perp;z) e^{ikz}
  = \frac{e^{ikz}}{i\lambda z} \int e^{ik\|{\bf r}_\perp - {\bf r}'_\perp \|^2/2z}\ \! \psi({\bf r}'_\perp;0) d{\bf r}'_\perp .
 \label{eqA9}
\ee



\section{Interference with Arbitrarily Polarized Beams}
\label{appB}

The~expression for the~total electric and~magnetic fields that arise from the~coherent superposition
of the~electric and~magnetic fields (\ref{eq65}) and~(\ref{eq66}) are
\begin{small}
\ba
 {\bf E}({\bf r}_\perp;z) & = & E_0
   \Bigg\{ \Big[ \alpha_1 \psi_1({\bf r}_\perp;z) + \alpha_2 \psi_2({\bf r}_\perp;z) \Big] {\bf \hat{x}}
   + \bigg[ \beta_1 \psi_1({\bf r}_\perp;z) + \beta_2 \psi_2({\bf r}_\perp;z) \bigg] {\bf \hat{y}}
 \nonumber \\ & &
   + \frac{e^{-i(\varphi_z + \pi/2)}}{2k\sigma_0\sigma_z}\
     \bigg[ \Big( \alpha_1 \Delta x_1 + \beta_1 \Delta y_1 \Big) \psi_1({\bf r}_\perp;z)
          + \Big( \alpha_2 \Delta x_2 + \beta_2 \Delta y_2 \Big) \psi_2({\bf r}_\perp;z) \bigg] {\bf \hat{z}} \Bigg\}
 \nonumber \\
 & = & E_0
   \Bigg\{ \Big[ \alpha_1 \psi_1({\bf r}_\perp;z) + \alpha_2 \psi_2({\bf r}_\perp;z) \Big] {\bf \hat{x}}
   + \bigg[ \beta_1 \psi_1({\bf r}_\perp;z) + \beta_2 \psi_2({\bf r}_\perp;z) \bigg] {\bf \hat{y}}
  \\ & &
   + \frac{e^{-i(\varphi_z + \pi/2)}}{2k\sigma_0\sigma_z}\
     \bigg[ x \Big( \alpha_1 \psi_1({\bf r}_\perp;z) + \alpha_2 \psi_2({\bf r}_\perp;z) \Big)
        - x_c \Big( \alpha_1 \psi_1({\bf r}_\perp;z) - \alpha_2 \psi_2({\bf r}_\perp;z) \Big)
 \nonumber \\ & & \qquad \qquad \qquad \qquad
          + y \Big( \beta_1 \psi_1({\bf r}_\perp;z) + \beta_2 \psi_1({\bf r}_\perp;z) \Big) \bigg] {\bf \hat{z}} \Bigg\} ,
 \nonumber \label{eq65b} \\
 \nonumber
 \ea
 \end{small}

 \begin{small}
 \ba
 {\bf H}({\bf r}_\perp;z) & = & H_0
   \Bigg\{ - \Big[ \beta_1 \psi_1({\bf r}_\perp;z) + \beta_2 \psi_2({\bf r}_\perp;z) \Big] {\bf \hat{x}}
   + \bigg[ \alpha_1 \psi_1({\bf r}_\perp;z) + \alpha_2 \psi_2({\bf r}_\perp;z) \bigg] {\bf \hat{y}}
 \nonumber \\ & &
   + \frac{e^{-i(\varphi_z + \pi/2)}}{2k\sigma_0\sigma_z}\
     \bigg[ \Big( \beta_1 \Delta x_1 - \alpha_1 \Delta y_1 \Big) \psi_1({\bf r}_\perp;z)
          + \Big( \beta_2 \Delta x_2 - \alpha_2 \Delta y_2 \Big) \psi_2({\bf r}_\perp;z) \bigg] {\bf \hat{z}} \Bigg\}
 \nonumber \\
 & = & H_0
   \Bigg\{ - \Big[ \beta_1 \psi_1({\bf r}_\perp;z) + \beta_2 \psi_2({\bf r}_\perp;z) \Big] {\bf \hat{x}}
   + \bigg[ \alpha_1 \psi_1({\bf r}_\perp;z) + \alpha_2 \psi_2({\bf r}_\perp;z) \bigg] {\bf \hat{y}}
 \\ & &
   + \frac{e^{-i(\varphi_z + \pi/2)}}{2k\sigma_0\sigma_z}\
     \bigg[ x \Big( \beta_1 \psi_1({\bf r}_\perp;z) + \beta_2 \psi_2({\bf r}_\perp;z) \Big)
        - x_c \Big( \beta_1 \psi_1({\bf r}_\perp;z) - \beta_2 \psi_2({\bf r}_\perp;z) \Big)
 \nonumber \\ & & \qquad \qquad \qquad \qquad
          - y \Big( \alpha_1 \psi_1({\bf r}_\perp;z) + \alpha_2 \psi_1({\bf r}_\perp;z) \Big) \bigg] {\bf \hat{z}} \Bigg\} ,
 \label{eq66b}
  \nonumber
\ea
\end{small}
from which, after some algebra, retaining terms up to the~lowest order in $\Delta x$ and~$\Delta y$, the
expression for the~intensity distribution, Equation~(\ref{density1b}), is obtained, with the~visibility factor
associated with the~polarization states of each beam as given by Equation~(\ref{overlap}).

By further proceeding with the~electric and~magnetic superposition fields, (\ref{eq65b}) and~(\ref{eq66b}),
we~can determine the~expression for the~time-averaged Poynting vector, with its components reading as
\begin{small}
\textls[-35]{\ba
 S_x/I_0 & = &
  \frac{z}{4k^2\sigma_0^2\sigma_z^2} \Bigg\{ \Delta x_1 \Big( |\alpha_1|^2 - |\beta_1|^2 \Big) |\psi_1|^2
                                          + \Delta x_2 \Big( |\alpha_2|^2 - |\beta_2|^2 \Big) |\psi_2|^2
          + 2 x {\rm Re} \bigg[ \Big( \alpha_1 \alpha_2^* - \beta_1 \beta_2^* \Big) \psi_1 \psi_2^* \bigg]
  \nonumber \\ & & \qquad \qquad
     + 2 y {\rm Re} \bigg[ \Big( \alpha_1 \beta_2^* + \alpha_2^* \beta_1 \Big) \psi_1 \psi_2^* \bigg]
     + 2 y {\rm Re} \Big( \alpha_1 \beta_1^* \Big) |\psi_1|^2
     + 2 y {\rm Re} \Big( \alpha_2 \beta_2^* \Big) |\psi_2|^2 \Bigg\}
   \\ & &
     + \frac{z_R x_c}{2k^2\sigma_0^2\sigma_z^2}\ {\rm Im} \bigg[ \Big( \alpha_1 \alpha_2^* - \beta_1 \beta_2^* \Big) \psi_1 \psi_2^* \bigg] ,
 \label{eq49bX} \nonumber\\
 S_y/I_0 & = &
  - \frac{z}{4k^2\sigma_0^2\sigma_z^2} \Bigg\{ y \Big( |\alpha_1|^2 - |\beta_1|^2 \Big) |\psi_1|^2
                                             + y \Big( |\alpha_2|^2 - |\beta_2|^2 \Big) |\psi_2|^2
          + 2 y {\rm Re} \bigg[ \Big( \alpha_1 \alpha_2^* - \beta_1 \beta_2^* \Big) \psi_1 \psi_2^* \bigg]
  \nonumber \\ & & \qquad \qquad
     - 2 x {\rm Re} \bigg[ \Big( \alpha_1 \beta_2^* + \alpha_2^* \beta_1 \Big) \psi_1 \psi_2^* \bigg]
     - 2 \Delta x_1 {\rm Re} \Big( \alpha_1 \beta_1^* \Big) |\psi_1|^2
     - 2 \Delta x_2 {\rm Re} \Big( \alpha_2 \beta_2^* \Big) |\psi_2|^2 \Bigg\}
   \\ & &
     + \frac{z_R x_c}{2k^2\sigma_0^2\sigma_z^2}\ {\rm Im} \bigg[ \Big( \alpha_1 \beta_2^* + \alpha_2^* \beta_1 \Big) \psi_1 \psi_2^* \bigg] ,
 \label{eq49bY} \nonumber\\
 S_z/I_0 & = & |\psi|^2 ,
 \label{eq49bZ}
\ea}
\end{small}
where $I_0 = E_0 H_0/2$ and~$|\psi|^2$ in (\ref{eq49bZ}) as given by (\ref{density1b}).
The~ray equations along the~$x$ and~$y$ directions in terms of the~longitudinal coordinate $z$ are
then obtained from these components, being
\begin{small}
\textls[-35]{\ba
 \frac{dx}{dz} & = &
  \frac{z}{4k^2\sigma_0^2\sigma_z^2 |\psi|^2} \Bigg\{ \Delta x_1 \Big( |\alpha_1|^2 - |\beta_1|^2 \Big) |\psi_1|^2
                                          + \Delta x_2 \Big( |\alpha_2|^2 - |\beta_2|^2 \Big) |\psi_2|^2
          + 2 x {\rm Re} \bigg[ \Big( \alpha_1 \alpha_2^* - \beta_1 \beta_2^* \Big) \psi_1 \psi_2^* \bigg]
  \nonumber \\ & & \qquad \qquad
     + 2 y {\rm Re} \bigg[ \Big( \alpha_1 \beta_2^* + \alpha_2^* \beta_1 \Big) \psi_1 \psi_2^* \bigg]
     + 2 y {\rm Re} \Big( \alpha_1 \beta_1^* \Big) |\psi_1|^2
     + 2 y {\rm Re} \Big( \alpha_2 \beta_2^* \Big) |\psi_2|^2 \Bigg\}
   \\ & &
     + \frac{z_R x_c}{2k^2\sigma_0^2\sigma_z^2 |\psi|^2}\ {\rm Im} \bigg[ \Big( \alpha_1 \alpha_2^* - \beta_1 \beta_2^* \Big) \psi_1 \psi_2^* \bigg] ,
 \label{eq56b} \nonumber\\
 \frac{dy}{dz} & = &
  - \frac{z}{4k^2\sigma_0^2\sigma_z^2 |\psi|^2} \Bigg\{ y \Big( |\alpha_1|^2 - |\beta_1|^2 \Big) |\psi_1|^2
                                             + y \Big( |\alpha_2|^2 - |\beta_2|^2 \Big) |\psi_2|^2
          + 2 y {\rm Re} \bigg[ \Big( \alpha_1 \alpha_2^* - \beta_1 \beta_2^* \Big) \psi_1 \psi_2^* \bigg]
  \nonumber \\ & & \qquad \qquad
     - 2 x {\rm Re} \bigg[ \Big( \alpha_1 \beta_2^* + \alpha_2^* \beta_1 \Big) \psi_1 \psi_2^* \bigg]
     - 2 \Delta x_1 {\rm Re} \Big( \alpha_1 \beta_1^* \Big) |\psi_1|^2
     - 2 \Delta x_2 {\rm Re} \Big( \alpha_2 \beta_2^* \Big) |\psi_2|^2 \Bigg\}
   \\ & &
     + \frac{z_R x_c}{2k^2\sigma_0^2\sigma_z^2 |\psi|^2}\ {\rm Im} \bigg[ \Big( \alpha_1 \beta_2^* + \alpha_2^* \beta_1 \Big) \psi_1 \psi_2^* \bigg] .
 \label{eq56bY}\nonumber
\ea}
\end{small}


\reftitle{References}

\end{document}